\documentclass[%
 reprint,
 amsmath,amssymb,
 aps,
prb,
]{revtex4-1}

\usepackage{graphicx}
\usepackage[caption=false]{subfig}
\usepackage{dcolumn}
\usepackage{bm}

\usepackage{setspace}
\usepackage{tabularx,ragged2e,booktabs}
\newcommand{\ra}[1]{\renewcommand{\arraystretch}{#1}}

\usepackage{lipsum}

\begin{document}

\preprint{APS/123-QED}

\title{A Robust Multi-Scale Field-Only Formulation of Electromagnetic Scattering}

\author{Qiang Sun}
\email{Qiang.Sun@unimelb.edu.au}
\affiliation{Particulate Fluids Processing Center, Department of Chemical and Biomolecular Engineering, The University of Melbourne, Parkville 3010 VIC, Australia
}

\author{Evert Klaseboer}
\email{evert@ihpc.a-star.edu.sg}
\affiliation{Institute of High Performance Computing, 1 Fusionopolis Way, Singapore 138632, Singapore}

\author{Derek Y. C. Chan}
\email{D.Chan@unimelb.edu.au}
\homepage{http://D.Chan.is}
\affiliation{Particulate Fluids Processing Center, School of Mathematics and Statistics, The University of Melbourne, Parkville 3010 VIC, Australia, and \\
Department of Mathematics, Swinburne University of Technology, Hawthorn 3122 VIC, Australia}

\date{\today}

\begin{abstract}
We present a boundary integral formulation of electromagnetic scattering by homogeneous bodies that are characterized by linear constitutive equations in the frequency domain. By working with the Cartesian components of the electric, $\boldsymbol{E}$ and magnetic, $\boldsymbol{H}$ fields and with the scalar functions $(\boldsymbol{r \cdot E})$ and $(\boldsymbol{r \cdot H})$ where $\boldsymbol{r}$ is a position vector, the problem can be cast as having to solve a set of scalar Helmholtz equations for the field components that are coupled by the usual electromagnetic boundary conditions at material boundaries. This facilitates a direct solution for the surface values of $\boldsymbol{E}$ and $\boldsymbol{H}$ rather than having to work with surface currents or surface charge densities as intermediate quantities in existing methods. Consequently, our formulation is free of the well-known numerical instability that occurs in the zero frequency or long wavelength limit in traditional surface integral solutions of Maxwell's equations and our numerical results converge uniformly to the static results in the long wavelength limit. Furthermore, we use a formulation of the scalar Helmholtz equation that is expressed as classically convergent integrals and does not require the evaluation of principal value integrals or any knowledge of the solid angle. Therefore, standard quadrature and higher order surface elements can readily be used to improve numerical precision for the same number of degrees of freedom. In addition, near and far field values can be calculated with equal precision and multiscale problems in which the scatterers possess characteristic length scales that are both large and small relative to the wavelength can be easily accommodated. From this we obtain results for the scattering and transmission of electromagnetic waves at dielectric boundaries that are valid for any ratio of  the local surface curvature to the wave number. This is a generalization of the familiar Fresnel formula and Snell's law, valid at planar dielectric boundaries, for the scattering and transmission of electromagnetic waves at surfaces of arbitrary curvature. Implementation details are illustrated with scattering by multiple perfect electric conductors as well as dielectric bodies with complex geometries and composition. 
\begin{description}
\item[PACS numbers]
42.25.Fx, 42.25.-p, 20.30.Rz
\end{description}
\end{abstract}

\pacs{Valid PACS appear here}
\maketitle


\section{\label{sec:intro} Introduction}

Accurate numerical solutions of Maxwell's partial differential equations that describe the propagation of electromagnetic waves~\cite{Maxwell_1865} are vital to a diverse range of applications ranging from radar telemetry, biomedical imaging, wireless communications to nano-photonics that collectively span length scales of around 12 orders of magnitude. At present there are two main approaches to solving Maxwell's equations that are distinguished by working either in the time domain or in the frequency domain. The first of which is the finite difference time domain (FDTD) method of Yee~\cite{Yee_1966} that solves the Maxwell's partial differential equations in space and time variables for the electric and magnetic fields. Spatial derivatives of field values at discrete locations in the 3D domain are approximated by finite difference and their time evolution is tracked by time-stepping. The other approach is to work in the frequency domain where solutions of the partial differential equations in the spatial coordinates are expressed as surface integral equations based on the Stratton-Chu potential theory formulation~\cite{StrattonChu_1939}. The unknowns in the surface integral equations are the surface current densities. After these are found, the relevant field quantities are then obtained by post-processing~\cite{PoggioMiller_1973}. A related approach formulates the problem in terms of surface charges and currents from which the electromagnetic vector and scalar potentials can be found and field quantities are then obtained by further differentiation~\cite{ChangHarrington_1977, Garcia_2002}.

Each of the time domain or frequency domain methods has its own advantages and challenges. With the time domain approach, one solves directly for the electric and magnetic fields. However, if working in an infinite spatial domain, it is necessary to account for the conditions at infinity numerically, such as the Sommerfeld radiation condition.  For problems having material boundaries with multiple characteristic length scales, special considerations have to be paid to constructing the 3D grid geometry to ensure sufficient accuracy when derivatives are approximated by finite differences.  For materials that exhibit frequency dispersion, the material constitutive equations appear in time convolution form~\cite{LandauLifshitz_1960, Schneider_2010} whereas experimental data for dielectric permitivities are generally measured and characterized in the frequency domain.

On the other hand, working in the frequency domain for problems involving linear dielectrics, the constitutive equations only involve material properties as constants of proportionality. The spatial solution can then be written as surface integrals~\cite{PoggioMiller_1973} whereby conditions at infinity can be accounted for analytically. The use of a surface integral formulation means the reduction of spatial dimensionality by one and geometric features of interfaces can be accommodated more readily.  Here, the unknowns to be determined are surface currents so that field quantities need to be found by subsequent post-processing. However, due to the use of the Green's function in the formulation, the surface integrals are divergent in the classical sense and need to be interpreted as principal value integrals~\cite{Yla_2014}, thus requiring extra effort in numerical implementations. A rather serious limiting feature of the surface integral approach is the so called `zero frequency catastrophe'~\cite{ChewBook_2009,Vico_2016} in which the surface integrals become numerically ill-conditioned in the limit when the wavelength becomes much larger than the characteristic dimension of the problem. Numerically, this arises because this formulation couples the electric and magnetic fields that become independent in the long wavelength or electrostatic limit.

In this paper, we develop a boundary integral formulation of the solution of Maxwell's equations for linear homogeneous dielectrics in the frequency domain using a conceptually simpler and numerically more robust approach. We seek to retain the advantages of the time domain method by working directly with the electric and magnetic fields, but without the need to solve for intermediate quantities such as surface currents and charges.  We retain the surface integral formulation that has the advantage of reduction of spatial dimension and the analytical account of conditions at infinity. By working with Cartesian components of the electric and magnetic fields, we use a recently developed boundary integral formulation of the solution of the scalar Helmholtz equation that is free of singularities in the integrands~\cite{Sun_2015b}. Such an approach reflects correctly the reality that the physical problem has no singularities at material boundaries and there is complete symmetry in finding the electric or magnetic fields. We can work separately with the electric and magnetic fields so that in the long wavelength limit, our formulation reduces naturally to the electrostatic and magnetostatic limits without numerical instabilities. In this way, our approach retains the advantages of the current time and frequency domain methods of solving the Maxwell's equations.

To put our formulation into context, we first provide in Sec.~\ref{sec:background}, a brief summary of the current state of the boundary integral equation solution of Maxwell's equations based on the Stratton-Chu formalism and to make explicit currently known challenges of this approach. In Sec.~\ref{sec:formulation} we show how the solution of Maxwell's equations can be cast only in terms of the Cartesian components of the electric, $\boldsymbol{E}$ and magnetic, $\boldsymbol{H}$ fields, and that these can be found from the solution of distinct sets of scalar Helmholtz equations. In Sec.~\ref{sec:NSBIE} we develop boundary integral solutions of the scalar Helmholtz equations for which the surface integrals do not have divergent integrands. To bring out the key features of our approach, we consider implementation of our formulation for the simpler case of scattering by perfect electrical conductors in Sec.~\ref{sec:PEC}.  As expected, scattering by dielectric bodies considered in Sec.~\ref{sec:DIEL} is more complex in technical details but the basic framework is the same. The key result in this case is a generalised Fresnel condition and Snell's law at a curved dielectric boundary that reduces naturally to the familiar results in the limit of a planar boundary. Finally in Sec.~\ref{sec:ZEROFREQ} we show that our field-only formulation does not suffer from the well-documented numerical instability at the zero frequency or long wavelength limit that is inherent in all current boundary integral formulation of electromagnetic scattering~\cite{ChewBook_2009, Vico_2016}, so that our results will reduce uniformly to the correct static limit. Thus our boundary integral formulation provides a theoretically and numerically robust resolution to the so called zero frequency catastrophe. We illustrate our approach by benchmarking against known problems that have analytic solutions and also consider examples with multiple scatterers and scatterers with layered geometries or with very different characteristic length scales.

\section{\label{sec:background} Stratton-Chu boundary integral formulation}

For propagation in homogeneous media, with linear constitutive equations for the displacement field, $\boldsymbol{D}$ and magnetic induction, $\boldsymbol{B}$:~\cite{Kerker_1969}
\begin{subequations} \label{eq:constitutive}
\begin{eqnarray}
\boldsymbol{D}(\boldsymbol{r}, \omega) &=& \epsilon \; \boldsymbol{E}(\boldsymbol{r}, \omega)    \label{eq:constitutive_a}
\\
\boldsymbol{B}(\boldsymbol{r}, \omega) &=& \mu \; \boldsymbol{H}(\boldsymbol{r}, \omega)         \label{eq:constitutive_b}
\end{eqnarray}
\end{subequations}
where $\epsilon = \epsilon_0 \; \epsilon_r(\omega)$ and $\mu = \mu_0 \; \mu_r(\omega)$ are the frequency dependent permittivity and permeability of the material, Maxwell's equations in the frequency domain with harmonic time dependence $\exp(-i \omega t)$ read~\cite{Maxwell_1865}:
\begin{subequations} \label{eq:Maxwell}
\begin{eqnarray}
\nabla \times \boldsymbol{E}&=& i \omega \; \mu \; \boldsymbol{H}      \label{eq:Maxwell_a}
\\
\nabla \times \boldsymbol{H}&=& - i \omega \; \epsilon \; \boldsymbol{E} +  \boldsymbol{j}      \label{eq:Maxwell_b}
\\
\nabla \cdot \boldsymbol{D}&=& \rho      \label{eq:Maxwell_c}
\\
\nabla \cdot \boldsymbol{B}&=& 0.      \label{eq:Maxwell_d}
\end{eqnarray}
\end{subequations}
In scattering problems, the imposed incident fields, ($\boldsymbol{E}^{inc}$, $\boldsymbol{H}^{inc}$) are specified instead of the sources represented by the real volume current density, $\boldsymbol{j}(\boldsymbol{r}, \omega)$ and charge density, $\rho(\boldsymbol{r}, \omega)$ in (\ref{eq:Maxwell}).

The Stratton-Chu formulation~\cite{StrattonChu_1939} uses potential theory and the vector Green's Theorem to express a formally exact solution of the Maxwell's equations in a volume domain in terms of integrals over values of the electric and magnetic fields on the boundaries enclosing the domain. The electric field, $\boldsymbol{E}(\boldsymbol{r}_0)$ and magnetic field, $\boldsymbol{H}(\boldsymbol{r}_0)$ at an observation point $\boldsymbol{r}_0$ in the 3D domain, $V$ are expressed in terms of the following integrals over the surface, $S$ enclosing $V$, with implicit frequency, $\omega$ dependence
\begin{subequations} \label{eq:EH_StrattonChu}
\begin{eqnarray}
\boldsymbol{E}(\boldsymbol{r}_0) = \boldsymbol{E}^{inc}(\boldsymbol{r}_0) 
- \frac{1}{c_0}\int_S [ i \omega \mu (\boldsymbol{n} \times \boldsymbol{H})G \qquad  \qquad  \nonumber \\
 + (\boldsymbol{n} \times \boldsymbol{E}) \times \nabla G + (\boldsymbol{n} \cdot \boldsymbol{E}) \nabla G ] \; dS(\boldsymbol{r})   \qquad     \label{eq:EH_StrattonChu_E}
\\
\nonumber \\
\boldsymbol{H}(\boldsymbol{r}_0) = \boldsymbol{H}^{inc}(\boldsymbol{r}_0) 
+ \frac{1}{c_0}\int_S [ i \omega \epsilon (\boldsymbol{n} \times \boldsymbol{E})G \qquad  \qquad  \nonumber \\
 - (\boldsymbol{n} \times \boldsymbol{H}) \times \nabla G - (\boldsymbol{n} \cdot \boldsymbol{H}) \nabla G ] \; dS(\boldsymbol{r})   \qquad    \label{eq:EH_StrattonChu_H}
\end{eqnarray}
\end{subequations}
where the Green's function
\begin{eqnarray} \label{eq:Greens_function}
  G(\boldsymbol{r}, \boldsymbol{r}_0) = \frac{\exp(ik|\boldsymbol{r} - \boldsymbol{r}_0|)}{|\boldsymbol{r} - \boldsymbol{r}_0|}
\end{eqnarray}
satisfies: $[\nabla^2 + k^2]G(\boldsymbol{r}, \boldsymbol{r}_0) = - 4\pi \delta(\boldsymbol{r} - \boldsymbol{r}_0)$, $k^2 \equiv \omega^2 \epsilon \mu$; and unit vector $\boldsymbol{n}$ on $S$ points out of the domain $V$.

When $\boldsymbol{r}_0$ lies wholly within $V$ and not on the surface, $S$, the constant $c_0$ is $4 \pi$ and the surface integrals are well-defined in the classical sense. The incident fields, $\boldsymbol{E}^{inc}$ and $\boldsymbol{H}^{inc}$ in (\ref{eq:EH_StrattonChu}) can either be prescribed as an incident wave or be determined from volume integrals over the given sources that are the real current density, $\boldsymbol{j}$ and charge density $\rho$ in Maxwell's equations (\ref{eq:Maxwell})~\cite{StrattonChu_1939}
\begin{subequations} \label{eq:EH_inc}
\begin{eqnarray}
\boldsymbol{E}^{inc}(\boldsymbol{r}_0) &=& \frac{1}{c_0}\int_V [ i \omega \mu \boldsymbol{j} G + (\rho / \epsilon) \nabla G] \; dV(\boldsymbol{r})   \label{eq:EH_inc_E}
 \\
\boldsymbol{H}^{inc}(\boldsymbol{r}_0) &=& \frac{1}{c_0}\int_V  [ \boldsymbol{j} \times \nabla G ]  \; dV(\boldsymbol{r}).      \label{eq:EH_inc_H}
\end{eqnarray}
\end{subequations}

The above formalism of Stratton-Chu was used originally to calculate far field diffraction patterns using  analytical approximations of the surface integrals in (\ref{eq:EH_StrattonChu}). However, by putting the observation point $\boldsymbol{r}_0$ onto the surface, $S$ in (\ref{eq:EH_StrattonChu}), one can obtain surface integral equations involving only values of $\boldsymbol{E}$ and $\boldsymbol{H}$ on the surface~\cite{PoggioMiller_1973} and the numerical solutions of such equations became feasible with the availability of computational capabilities some two decades later~\cite{Jaswon_1963, Symm_1963}.
However, because both $\boldsymbol{r}$ and $\boldsymbol{r}_0$ are now on $S$, the nature of the singularity of $\nabla G$ in the integrands when $\boldsymbol{r} \rightarrow \boldsymbol{r}_0$ means that the surface integrals are divergent in the classical sense and so they must be interpreted as principal value integrals (denoted by $PV$) as in the treatment of generalized functions. Assuming the surface at $\boldsymbol{r}_0$ has a well-defined tangent plane, the constant $c_0$ becomes $2 \pi$, and we obtain the following principal value (PV) surface integral equations,~\cite{PoggioMiller_1973}
\begin{subequations} \label{eq:EH_Vector_BIE}
\begin{eqnarray}
\boldsymbol{E}(\boldsymbol{r}_0) = \boldsymbol{E}^{inc}(\boldsymbol{r}_0) 
- \frac{1}{c_0}\int_{S,PV} [ i \omega \mu (\boldsymbol{n} \times \boldsymbol{H})G \qquad  \qquad  \nonumber \\
 + (\boldsymbol{n} \times \boldsymbol{E}) \times \nabla G + (\boldsymbol{n} \cdot \boldsymbol{E}) \nabla G ] \; dS(\boldsymbol{r})   \qquad     \label{eq:EH_Vector_BIE_E}
\\
\nonumber \\
\boldsymbol{H}(\boldsymbol{r}_0) = \boldsymbol{H}^{inc}(\boldsymbol{r}_0) 
+ \frac{1}{c_0}\int_{S,PV} [ i \omega \epsilon (\boldsymbol{n} \times \boldsymbol{E})G \qquad  \qquad  \nonumber \\
 - (\boldsymbol{n} \times \boldsymbol{H}) \times \nabla G - (\boldsymbol{n} \cdot \boldsymbol{H}) \nabla G ] \; dS(\boldsymbol{r}).   \qquad    \label{eq:EH_Vector_BIE_H}
\end{eqnarray}
\end{subequations}

Although (\ref{eq:EH_Vector_BIE}) involves only $\boldsymbol{E}$ and $\boldsymbol{H}$ on the surface, their (numerical) solution is cast in terms of induced surface currents and surface charge densities from which $\boldsymbol{E}$ and $\boldsymbol{H}$ are determined by post-processing. To illustrate this, consider the simpler example of scattering by a perfect electrical conductor (PEC).

On the surface of a PEC, we have the condition that the tangential components of $\boldsymbol{E}$ and the normal component of $\boldsymbol{H}$ all vanish, that is, $\boldsymbol{n} \times \boldsymbol{E} = \boldsymbol{0}$ and $\boldsymbol{n} \cdot \boldsymbol{H} = 0$ on $S$ in the integrands of (\ref{eq:EH_Vector_BIE}). Furthermore, the induced surface current density is~\cite{StrattonChu_1939} $\boldsymbol{J} = \boldsymbol{H}  \times \boldsymbol{n}$ and is related to the induced surface charge density $\sigma = - \epsilon (\boldsymbol{n} \cdot \boldsymbol{E})$ by the continuity equation $\nabla \cdot \boldsymbol{J} = i \omega \sigma$. Thus the tangential components of (\ref{eq:EH_Vector_BIE}) have the form 
\begin{subequations} \label{eq:EH_PEC}
\begin{eqnarray}
\boldsymbol{n} \times \boldsymbol{E}^{inc}(\boldsymbol{r}_0) = \qquad \qquad \qquad \qquad \qquad \qquad \qquad  \qquad \quad  \nonumber \\
- \frac{1}{c_0}  \boldsymbol{n} \times \int_{S,PV} [ i \omega \mu \boldsymbol{J}G + \frac{1}{i \omega \epsilon}(\nabla \cdot \boldsymbol{J}) \nabla G ] \; dS(\boldsymbol{r}) \quad   \quad  \quad   \label{eq:EH_PEC_E}
\\
\nonumber \\
\boldsymbol{J}(\boldsymbol{r}_0) = - \boldsymbol{n} \times \boldsymbol{H}^{inc}(\boldsymbol{r}_0) \qquad \qquad \qquad \qquad \qquad \qquad \quad \nonumber \\
- \frac{1}{c_0}\boldsymbol{n} \times \int_{S,PV} \boldsymbol{J} \times \nabla G  \; dS(\boldsymbol{r}).   \qquad  \quad  \label{eq:EH_PEC_H}
\end{eqnarray}
\end{subequations}
For given incident fields $\boldsymbol{E}^{inc}$ and $\boldsymbol{H}^{inc}$, the induced surface current density, $\boldsymbol{J}$ can now be found by solving either of (\ref{eq:EH_PEC}) or their linear combination and then $\boldsymbol{E}$ and $\boldsymbol{H}$ can be obtained, for example, from the Stratton-Chu formula (\ref{eq:EH_StrattonChu}) by post-processing.

In practical applications, the surface $S$ is almost always represented by a tessellation of planar triangular elements and the unknowns are taken to be constants within each element so that the value of $c_0$ can be taken to be $2 \pi$. The use of higher order elements will generally require the calculation of $c_0$ that varies with the local surface geometry. At the element that contains both $\boldsymbol{r}$ and $\boldsymbol{r}_0$ additional steps are needed to evaluate the principal value integral.

A troubling feature of (\ref{eq:EH_PEC_E}) is that in the zero frequency limit, $\omega \rightarrow 0$ (or $k \rightarrow 0$), the second integrand becomes numerically unstable since $\nabla \cdot \boldsymbol{J}$ also vanishes in this limit~\cite{Vico_2016}. Thus this formulation is not robust in the multiscale sense if the characteristic length scale of the problem becomes much smaller than the wavelength. This is an unsurprising result because in the long wavelength or electrostatic limit, the electric and magnetic fields decouple and the electric field should be described in terms of the charge density rather than the current density~\cite{Vico_2016}. Furthermore, if the geometry of the problem has surfaces that are close together relative to the wavelength, the near singular behavior on one surface can have an adverse impact on the numerical precision of integrals over proximal surfaces. 

The result in (\ref{eq:EH_PEC_E}) has also been cast into a more compact form using the dyadic Green's function~\cite{Yaghjian_1980}
\begin{eqnarray} \label{eq:E_dyadic}
\boldsymbol{n} \times \boldsymbol{E}^{inc}(\boldsymbol{r}_0) = \qquad \qquad \qquad \qquad \qquad \qquad \qquad  \qquad \nonumber \\
- \frac{1}{c_0}  \boldsymbol{n} \times \int_{S,PV}  i \omega \mu \boldsymbol{J} \;[\textbf{I} + (1/k^2) \nabla \nabla]G   \; dS(\boldsymbol{r})   \qquad
\end{eqnarray}
that now results in a stronger hypersingularity in the integrand at $\boldsymbol{r} = \boldsymbol{r}_0$. This is due to interchanging the order of differentiation and integration of an integral that is not absolutely convergent~\cite{Chew_1989} although there are established methods to regularize the hypersingularity~\cite{Contopanagos_2002} and to address the zero frequency limit~\cite{Zhao_1989}.

In spite of all the above reservations, the boundary integral method of solving Maxwell's equations has clear advantages. The numerical problem of having to find unknowns on boundaries reduces a 3D problem to 2D, and if the boundary has axial symmetry, it can be simplified further to line integrals after Fourier decomposition. Thus apart from the obvious savings in the reduction of dimension, for problems with complex surface geometries or those with multiple characteristic length scales, the problem of having to construct multiscale 3D grids can be avoided. The dense linear systems that arise from boundary integral equations can be handled using $O(N \log N)$ fast solvers~\cite{Engheta_1992}. 

Nonetheless, from a physical perspective, there are some fundamental theoretical issues with the above boundary integral formulation of solutions of Maxwell's equations in terms of surface currents:
\begin{itemize}
 \item The mathematical singularities in the boundary integral equations arise from the application of the vector Green's Theorem in the Stratton-Chu formulation that uses the free space Green's function (\ref{eq:Greens_function}). In actual physical problems, the field quantities are perfectly well-behaved on boundaries so a mathematical description that injects inherent singularities that are not present in the physical problem suggests that a more optimal representation of the physics of the problem should be sought.
\item The singular nature of the surface integral equations makes it difficult to obtain field values at or close to boundaries with high accuracy.  For applications that require precise determination of the surface field such as in quantifying geometric field enhancement effects in micro-photonics or in determining the radiation pressure by integrating the Maxwell stress tensor, such limitations in numerical precision are undesirable.
\item In problems where different parts of surfaces can be close together, for example, in an array of scatterers that are nearly in contact, the near singular behavior of proximal points on different surfaces will adversely affect or limit the achievable numerical precision. This is related directly to the numerical instability of the formulation in the zero frequency ($\omega \rightarrow 0$) or the long wavelength electrostatic limit ($k \rightarrow 0 $) limit. 
\item  In applying the Stratton-Chu formulation, there is the need to first solve for the induced surface currents from which the $\boldsymbol{E}$ and $\boldsymbol{H}$ fields are then obtained by post-processing. It may be more efficient if one can solve for the fields directly.
\end{itemize}

Here, motivated by the desire to circumvent the above inherent and somewhat limiting characteristics of existing approaches of solving Maxwell's equations by the boundary integral equation method, we develop a boundary integral formulation for the solution of Maxwell's equations that does not use surface charge or current densities as intermediate quantities. 

\section{\label{sec:formulation} Field-Only Formulation}

Our objective is to derive surface integral equations for the Cartesian components of the fields $\boldsymbol{E}$ and $\boldsymbol{H}$ that are solution of scalar Helmholtz equations. The equations for $\boldsymbol{E}$ and $\boldsymbol{H}$ are not coupled. Furthermore, we use a recently developed boundary integral formulation in which all surface integrals have singularity-free integrands~\cite{Klaseboer_2012, Sun_2015b} and the term involving the solid angle has been eliminated. The points of difference and consequent advantages of the method are: 
\begin {enumerate}
\item Components of $\boldsymbol{E}$ and $\boldsymbol{H}$ are computed directly at boundaries without the need for post-processing or having to handle singular or hypersingular integral equations. This imparts high precision in calculating, for example, surface field enhancement effects and the Maxwell stress tensor.
\item The elimination of the solid angle term facilitates the use of quadratic surface elements to represent the surface, $S$ more accurately and the use of appropriate interpolation to represent variations of functions within each element to a consistent level of precision. This enables the boundary integrals to be evaluated using standard quadrature to confer high numerical accuracy with fewer degrees of freedom.
\item The absence of singular integrands means that geometric configurations in which different parts of the boundary are very close together will not cause numerical instabilities, thus fields and forces between surfaces can be found accurately even at small separations when surfaces are in near contact.
\item The simplicity of the formulation in not requiring complex algorithms to handle singularities and principal value integrals means significant savings in coding effort and reduction of opportunities for coding errors.
\item Multiple domains connected by boundary conditions can be implemented with relative ease.
\item The accuracy of the numerical implementation means that the effects of any resonant solutions of the Helmholtz equation are small unless the wavenumber is extremely close to one of the resonant values, so that the resonant solution is not likely to affect practical applications if the present approach is used.
\end {enumerate}

The motivation for our field-only formulation is drawn from the celebrated~\cite{Mishchenko_2008} exact analytical solution of the Maxwell's equations for the Mie~\cite{Mie_1908} problem of the scattering of an incident plane wave by a sphere. A more familiar exposition of this solution due to Debye~\cite{Debye_1909} has since been reproduced in more accessible forms~\cite{vandeHulst_1957, Liou_1977}. 

In a source free region, the electric and magnetic fields $\boldsymbol{E}$ and $\boldsymbol{H}$ are divergence free 
\begin{subequations} \label{eq:div_EH_zero}
\begin{eqnarray}
\nabla \cdot \boldsymbol{E} &=& 0   \label{eq:div_EH_zero_E}
\\
\nabla \cdot \boldsymbol{H} &=& 0   \label{eq:div_EH_zero_H}
\end{eqnarray}
\end{subequations}
and satisfy the wave equation
\begin{subequations} \label{eq:EH_wave_eqn}
\begin{eqnarray}
\nabla^2 \boldsymbol{E} + k^2 \boldsymbol{E} &=& \boldsymbol{0}     \label{eq:EH_wave_eqn_a}
\\
\nabla^2 \boldsymbol{H} + k^2 \boldsymbol{H} &=& \boldsymbol{0}  \label{eq:EH_wave_eqn_b}
\end{eqnarray}
\end{subequations}
In curvilinear coordinate systems, the Laplacian of a vector function $\boldsymbol{A}$, $\nabla^2 \boldsymbol{A} \equiv [\nabla (\nabla \cdot \boldsymbol{A}) - \nabla \times (\nabla \times \boldsymbol{A}) ]$, couples different orthogonal components where $\boldsymbol{A}$ can be $\boldsymbol{E}$ or $\boldsymbol{H}$ in (\ref{eq:EH_wave_eqn}). Only in the Cartesian coordinate system does (\ref{eq:EH_wave_eqn}) separate into scalar Helmholtz equations for the individual components. 

The key feature of the Debye solution is to represent the solution of the vector wave equations for $\boldsymbol{E}$ and $\boldsymbol{H}$, in terms of a pair of scalar functions $\psi_E$ and $\psi_M$ known as Debye potentials~\cite{Kerker_1969} that satisfy the scalar Helmholtz wave equation
\begin{eqnarray}
  \nabla^2 \psi_{E,M}(\boldsymbol{r}) + k^2 \psi_{E,M}(\boldsymbol{r}) = 0.
\end{eqnarray}
The electric and magnetic fields can then be expressed in terms of the Debye potentials using the differential operator $\textbf{L} \equiv i (\boldsymbol{r} \times \nabla)$ where $\boldsymbol{r}=(x,y,z)$ is the position vector~\cite{Low_1997}
\begin{subequations} \label{eq:EH_psi}
\begin{eqnarray}
\boldsymbol{E}&=&\textbf{L} \psi_M  - \frac{i}{k} \nabla \times  \textbf{L} \psi_E     \label{eq:EH_psi_a}
\\
\boldsymbol{H}&=& - \frac{i}{k} \nabla \times  \textbf{L} \psi_M - \textbf{L} \psi_E .   \label{eq:EH_psi_b}
\end{eqnarray}
\end{subequations}
Although analytic solutions for the Debye potentials, $\psi_E$ and $\psi_M$ can be expressed as infinite series in spherical harmonics and spherical Bessel functions with unknown expansion coefficients, their relationship to the fields in (\ref{eq:EH_psi}) means that the equations that determine the coefficients by imposing boundary conditions on $\boldsymbol{E}$ and $\boldsymbol{H}$ are tractable only if the boundary surfaces are spheres. For systems of multiple spheres, multi-center expansions and addition theorems for the spherical harmonics and spherical Bessel functions need to be used to construct a solution~\cite{Garcia_1999}.
  
 Since the fields $\boldsymbol{E}$ and $\boldsymbol{H}$ satisfy the wave equation (\ref{eq:EH_wave_eqn}), we use the identity: $\nabla^2(\boldsymbol{r} \cdot \boldsymbol{V}) \equiv 2 (\nabla \cdot \boldsymbol{V}) + \boldsymbol{r} \cdot (\nabla^2 \boldsymbol{V})$ for a differentiable vector field, $\boldsymbol{V}$ to replace the divergence free conditions of $\boldsymbol{E}$ and $\boldsymbol{H}$ in (\ref{eq:div_EH_zero}) in a source free region by Helmholtz equations for the scalar functions $(\boldsymbol{r \cdot E})$ and $(\boldsymbol{r \cdot H})$ as:
\begin{subequations} \label{eq:div_EH}
\begin{eqnarray}
2(\nabla \boldsymbol{\cdot E}) &\equiv& \nabla^2 (\boldsymbol{r \cdot E})+ k^2 (\boldsymbol{r \cdot E}) = 0     \label{eq:div_EH_a}
\\
2(\nabla \boldsymbol{\cdot H}) &\equiv& \nabla^2 (\boldsymbol{r \cdot H}) + k^2 (\boldsymbol{r \cdot H}) = 0.   \label{eq:div_EH_b}
\end{eqnarray}
\end{subequations}
It is easy to verify with the addition of a constant vector to $\boldsymbol{r}$ that this identity is independent of the choice of the origin of the coordinate system.

The results (\ref{eq:EH_wave_eqn}) and (\ref{eq:div_EH}) appear to be first demonstrated explicitly by Lamb~\cite{Lamb_1881} for elastic vibrations, but the significance here is that the Cartesian components of $\boldsymbol{E}$ and $\boldsymbol{H}$ and the scalar functions $(\boldsymbol{r \cdot E})$ and $(\boldsymbol{r \cdot H})$ all satisfy the scalar Helmholtz equation and are coupled by the continuity of the tangential components of $\boldsymbol{E}$ and $\boldsymbol{H}$ across material boundaries.

Therefore, the solution for $\boldsymbol{E}$ or for $\boldsymbol{H}$ can each be represented as a coupled set of 4 scalar equations:
\begin{eqnarray} \label{eq:Helm_p}
  \nabla^2 p_i(\boldsymbol{r}) + k^2 p_i(\boldsymbol{r}) = 0,  \qquad i = 1 .. 4
 \end{eqnarray} 
 where the scalar function $p_i(\boldsymbol{r})$, $i=1..4$, denotes $(\boldsymbol{r \cdot E})$ or one of the 3 Cartesian components of $\boldsymbol{E}$  for the electric field or $(\boldsymbol{r \cdot H}$) and $\boldsymbol{H}$ for the magnetic field. Thus the equations for $\boldsymbol{E}$ and for $\boldsymbol{H}$ can be solved separately.
 
 In general, the boundary integral representation of the solution of (\ref{eq:Helm_p}) expresses the solution for $p_i(\boldsymbol{r}_0)$ in the 3D solution domain in terms of an integral involving $p_i(\boldsymbol{r})$ and its normal derivative $\partial p_i(\boldsymbol{r}) / \partial n \equiv \boldsymbol{n \cdot} \nabla p_i(\boldsymbol{r})$ on the surface, $S$ that encloses the solution domain with outward unit normal $\boldsymbol{n}$. By putting $\boldsymbol{r}_0$ onto the surface and using a given boundary condition on $p_i(\boldsymbol{r})$ or on $\partial p_i(\boldsymbol{r}) / \partial n$, we obtain a surface integral equation to be solved~\cite{Sun_2015b}.

Consider first the equations involving the electric field quantities $\boldsymbol{E}$ and $(\boldsymbol{r \cdot E}$) on the surface, $S$. The vector wave equation (\ref{eq:EH_wave_eqn_a})  furnishes 3 relations between 6 unknowns, namely, $E_{\alpha}$ and $\partial E_{\alpha} / \partial n$, $(\alpha =x,y,z)$ and the relation (\ref{eq:div_EH_a}) between $(\boldsymbol{r} \cdot \boldsymbol{E})$ and $\partial (\boldsymbol{r} \cdot \boldsymbol{E}) / \partial n$ gives one more relation between $E_{\alpha}$ and $\partial E_{\alpha} / \partial n$ by using the identity: $\partial (\boldsymbol{r} \cdot \boldsymbol{E}) / \partial n = \boldsymbol{n} \cdot \boldsymbol{E} \; + \; \boldsymbol{r} \cdot \partial \boldsymbol{E} / \partial n$. The electromagnetic boundary conditions on the continuity of the tangential components of $\boldsymbol{E}$ provide the remaining 2 equations that then allows $\boldsymbol{E}$ and $\partial \boldsymbol{E} / \partial n$ on the surface to be determined.

A similar consideration also applies to the magnetic field quantities $\boldsymbol{H}$ and $(\boldsymbol{r \cdot H}$).

With the present field-only formulation, the Cartesian components of the electric field, $\boldsymbol{E}$ and the magnetic field, $\boldsymbol{H}$ are determined separately by a similar set of coupled scalar Helmholtz equations. In fact, the governing equations (\ref{eq:EH_wave_eqn}) and (\ref{eq:div_EH}) are identical with the interchange of $\boldsymbol{E}$ and $\boldsymbol{H}$ and the permittivity, $\epsilon$ and permeability, $\mu$. In the zero frequency ($\omega \rightarrow 0$) or long wavelength electrostatic limit ($k \rightarrow 0$), the two sets of Helmholtz equations simply reduce to Laplace equations and no numerical instability arises. Consequently, in multiscale problems with different characteristic lengths, $a_i$, this formulation will be stable against any variation in the range of the non-dimensional scaling parameters, $k a_i$, unlike the traditional formulation involving surface currents as, for example, in (\ref{eq:EH_PEC}) or (\ref{eq:E_dyadic}).

Before we discuss the solution of (\ref{eq:EH_wave_eqn}) and (\ref{eq:div_EH}) for scattering by perfect electrical conductors and dielectric bodies in Secs.\ref{sec:PEC} and \ref{sec:DIEL} respectively, we first consider the solution of the scalar Helmholtz equation (\ref{eq:Helm_p}) using a boundary integral formulation that does not contain any singularities in the surface integrals.

\section{\label{sec:NSBIE} Non-singular boundary integral solution of Helmholtz equation}

The conventional boundary integral solution of the scalar Helmholtz equation (\ref{eq:Helm_p}) is constructed from Green's Second Identity that gives an integral relation between $p(\boldsymbol{r})$ and its normal derivative $\partial{p}/\partial{n}$ (suppressing the subscript $i$) at points $\boldsymbol{r}$ and $\boldsymbol{r}_0$, both located on the boundary, $S$.   Such a solution of~(\ref{eq:Helm_p}) involves integrals of the Green's function, $G(\boldsymbol{r},\boldsymbol{r}_0)$ in (\ref{eq:Greens_function}) and its normal derivative $\partial G / \partial n$~\cite{becker1992}
\begin{eqnarray} \label{eq:CBIE}
c_0 p(\boldsymbol{r}_0) +\int_{S}^{} {p(\boldsymbol{r}) \frac{\partial {G}} {{\partial {n}}} \; dS(\boldsymbol{r}}) =   \int_{S}^{} {\frac{\partial {p (\boldsymbol{r})}} {{\partial {n}}} G  \; dS(\boldsymbol{r}}) \quad
\end{eqnarray} 
where $c_0$ is the solid angle at $\boldsymbol{r}_0$. The use of the Green's function $G$ means that the radiation condition for the scattered field at infinity can be satisfied exactly. Although both $G$ and $\partial G / \partial n$ are divergent at $\boldsymbol{r}_0 = \boldsymbol{r}$, the integrals are in fact integrable in the classical sense although extra effort is required to handle such integrable singularities in numerical solutions of (\ref{eq:CBIE}).

It turns out that it is possible to remove analytically all such integrable singular behavior associated with $G$ and $\partial G / \partial n$. This is accomplished by first constructing the conventional boundary integral equation as in (\ref{eq:CBIE}) for a related problem. Then by subtracting this from the original boundary integral equation gives an integral equation that does not contain any singularities in the integrands. In the process, the term involving the solid angle, $c_0$ has also been eliminated. This approach is called the Boundary Regularized Integral Equation Formulation (BRIEF)~\cite{Sun_2015b} and has been applied to problems in fluid mechanics and elasticity~\cite{Klaseboer_2012, Sun_2013, Sun_2015a}, colloidal and molecular electrostatics \cite{Sun_2016} and in solving the Laplace equation~\cite{Sun_2014}. Not having to handle such singularities confers simplifications in implementing numerical solutions with the additional flexibility to use higher order surface elements that can represent the surface geometry more accurately. The absence of singular integrands and the solid angle also means that it is easy to use accurate quadrature and interpolation methods to evaluate the surface integrals that will result in an increase in precision for the same number of degrees of freedom~\cite{Sun_2015b}. 

We start by considering the corresponding boundary integral equation for a function, $\phi(\boldsymbol{r})$ that is constructed to depend on the value of $p$ and $\partial p / \partial n$ at $\boldsymbol{r}_0$ as follows~\cite{Klaseboer_2012, Sun_2015b}
\begin{eqnarray} \label{eq:phi}
\phi(\boldsymbol{r}) \equiv p(\boldsymbol{r}_0) g(\boldsymbol{r}) + 
 \frac{\partial {p (\boldsymbol{r}_0)}} {{\partial {n}}}  f(\boldsymbol{r}).
\end{eqnarray} 
The functions $f(\boldsymbol{r})$ and $g(\boldsymbol{r})$ are chosen to satisfy the Helmholtz equation. Without loss of generality, we can also require $f(\boldsymbol{r})$ and $g(\boldsymbol{r})$  to satisfy the following conditions at $\boldsymbol{r} =\boldsymbol{r}_0$:
\begin{subequations} \label{eq:fg_constraints}
\begin{eqnarray}
f(\boldsymbol{r}_0) &=& 0,  \qquad   \boldsymbol{n}(\boldsymbol{r}_0) \cdot  \nabla f(\boldsymbol{r}_0) = 1
\\
g(\boldsymbol{r}_0) &=& 1, \qquad \boldsymbol{n}(\boldsymbol{r}_0) \cdot \nabla g(\boldsymbol{r}_0) = 0.
\end{eqnarray}
\end{subequations}
 There are many possible and convenient choices of $f(\boldsymbol{r})$ and $g(\boldsymbol{r})$ that satisfy (\ref{eq:fg_constraints})~\cite{Klaseboer_2012,Sun_2015b}.

The singularities and the solid angle, $c_0$ in~(\ref{eq:CBIE}) can now be removed by subtracting~(\ref{eq:CBIE}) from the corresponding boundary integral equation for $\phi(\boldsymbol{r})$ that is defined in~(\ref{eq:phi}), to give
\begin{eqnarray} \label{eq:NS_BIE}
\int_{S}^{} {[p(\boldsymbol{r}) - p(\boldsymbol{r}_0) g(\boldsymbol{r}) - \frac{\partial {p(\boldsymbol{r}_0)}} {{\partial {n}}}  f(\boldsymbol{r})] \frac{\partial {G}} {{\partial {n}}} dS(\boldsymbol{r}}) = \qquad \nonumber \\  \int_{S}^{} {G [\frac{\partial {p (\boldsymbol{r})}} {{\partial {n}}} - p(\boldsymbol{r}_0) \frac{\partial {g (\boldsymbol{r})}} {{\partial {n}}} - \frac{\partial {p(\boldsymbol{r}_0)}} {{\partial {n}}} \frac{\partial {f (\boldsymbol{r})}} {{\partial {n}}} ] dS(\boldsymbol{r}}). \quad 
\end{eqnarray} 
This is the key result of  the Boundary Regularized Integral Equation Formulation (BRIEF)~\cite{Sun_2015b} for the scalar Helmholtz equation. For example, if $p$ (Dirichlet) or $\partial{p}/\partial{n}$ (Neumann) is known, then (\ref{eq:NS_BIE}) can be solved for $\partial{p}/\partial{n}$ or $p$, respectively. 

With the properties of $f(\boldsymbol{r})$ and $g(\boldsymbol{r})$ given in (\ref{eq:fg_constraints}), the terms that multiply $G$ and $\partial G/\partial n$ vanish at the same rate as the rate of divergence of $G$ or $\partial G/\partial n$ as $\boldsymbol{r} \rightarrow \boldsymbol{r}_0$~\cite{Klaseboer_2012, Sun_2015b} and consequently both integrals in (\ref{eq:NS_BIE}) have non-singular integrands. The absence of the solid angle, $c_0$ in (\ref{eq:NS_BIE}) means the surface, $S$ can be represented to higher precision using quadratic elements with nodes at the vertices and boundaries of such elements at which we compute values of  $p$ and $\partial{p}/\partial{n}$ without having to calculate the solid angle at each node. Since the integrands are non-singular, we can represent variations in $p$ or $\partial{p}/\partial{n}$ within each surface element by interpolation between the node values and thus we can evaluate the surface integral more accurately by simple quadrature for all elements, including the one that contains both $\boldsymbol{r}$ and $\boldsymbol{r}_0$. 

In contrast, with the conventional boundary integral formulation in (\ref{eq:CBIE}), special treatment is necessary to perform the numerical integration over the element that contains the observation point $\boldsymbol{r}_0$ due to the divergence in the integrand. It is also common in the conventional approach to use only planar elements to represent the boundary, $S$ and to assume the functions $p$ and $\partial{p}/\partial{n}$ are constant over each element so that one is able to use $c_0 = 2 \pi$ as the solid angle. Otherwise, if node values at the vertices of such triangles are used as unknowns, the solid angle $c_0$ will no longer be $2\pi$, but its value at each node has to be computed from the local geometry.

Once the surface field values have been obtained, $\boldsymbol{E}$ and $\boldsymbol{H}$ at position, $\boldsymbol{r}_0$ anywhere in the solution domain or on the surface can be obtained by a numerically robust method~\cite{Sun_2015b}.

\section{\label{sec:PEC} Perfect electrical conductor (PEC) scatterers}

\subsection{\label{sec:PEC_formulation} PEC formulation}

We now give details of implementing our field-only formulation of electromagnetic scattering given by (\ref{eq:EH_wave_eqn}) and (\ref{eq:div_EH}) for the simpler case of scattering by a perfect electrical conductor (PEC) for which only the scattered field needs to be found~\cite{Klaseboer_2017}. The objective is to show explicitly how the solutions of the scalar Helmholtz equations for the Cartesian components of $\boldsymbol{E}$ and $(\boldsymbol{r \cdot E}$) or of $\boldsymbol{H}$ and $(\boldsymbol{r \cdot H}$) are determined by the electromagnetic boundary condition on the tangential components of $\boldsymbol{E}$ and the normal component of $\boldsymbol{H}$ at the PEC surface. This problem is simpler than scattering by dielectric bodies that will be considered in the section that follows, where both the scattered and transmitted fields need to be found.

We first consider the solution for the electric field. On the surface of a PEC, the tangential components of the total electric field, $\boldsymbol{E}$ vanish so it is convenient to work in terms of the normal component, $E_n = \boldsymbol{n} \cdot \boldsymbol{E}$ of the electric field at the surface. Physically, $E_n$ is proportional to the induced surface charge density on the PEC. 

For scattering by a PEC, the solution for the electric field is determined by the value of 4 scalar functions on the surface, namely: $\partial E_x/ \partial n, \partial E_y/ \partial n, \partial E_z/ \partial n$ and $E_n$. We decompose $\boldsymbol{E}$ into a sum of the incident field, $\boldsymbol{E}^{inc}$ and the scattered field, $\boldsymbol{E}^{scat}$ and solve for the above quantities for the scattered field using the boundary conditions that on the surface of the PEC, the tangential components of the scattered field cancel those of the incident field. The number of unknowns to be found is the same as for the classic solution of the scattering problem by a PEC sphere using a pair of scalar Debye potentials, $\psi_E$ and $\psi_M$, in which the 2 functions and their normal derivatives have to be found~\cite{vandeHulst_1957, Liou_1977}. 

The formulation for the magnetic field, $\boldsymbol{H}$ is similar, but at PEC boundaries, (\ref{eq:div_EH_b}) is equivalent to the boundary condition that the normal component of the total magnetic field $\boldsymbol{H}$ vanishes on the PEC:
\begin{eqnarray} \label{eq:Hn_eq_0}
     \boldsymbol{n \cdot H} = 0 \quad \text{on} \quad S
\end{eqnarray}
so the tangential components of $\boldsymbol{H}$ are the unknowns to be found. They can be determined from the boundary condition on the tangential component of $\boldsymbol{E}$ as follows. For a chosen unit tangent $\boldsymbol{t}_1$ on the surface, the orthogonal unit tangent is $\boldsymbol{t}_2 \equiv \boldsymbol{t}_1 \times \boldsymbol{n}$ on $S$. Then using Ampere's law, we express the component of  $\boldsymbol{E}$ parallel to $\boldsymbol{t}_2$, namely, $E_{t_2} \equiv \boldsymbol{E} \boldsymbol{\cdot} \boldsymbol{t}_2 = \boldsymbol{E} \cdot (\boldsymbol{t}_1 \times \boldsymbol{n}) = \boldsymbol{t}_1 \cdot (\boldsymbol{n} \times \boldsymbol{E})$, in terms of $\boldsymbol{H}$
\begin{subequations} \label{eq:Ep_to_H}
\begin{eqnarray} 
  E_{t_2} &=& \boldsymbol{t}_1 \cdot (\boldsymbol{n} \times \boldsymbol{E})  = \frac{i}{ \omega \epsilon} \{ \boldsymbol{t}_1 \cdot (\boldsymbol{n} \times \nabla \times \boldsymbol{H}) \} \label{eq:Et} 
   \\
         &=& \frac{i}{ \omega \epsilon} \{ \boldsymbol{n} \cdot (\boldsymbol{t}_1 \cdot \nabla) \boldsymbol{H} - \boldsymbol{t}_1 \cdot (\boldsymbol{n} \cdot \nabla) \boldsymbol{H} \}  = 0. \label{eq:Et_EQ_0} 
\end{eqnarray}
\end{subequations}
The second equality in (\ref{eq:Et_EQ_0}) follows from the condition that the tangential component of the electric field vanishes on the PEC surface, $S$. By choosing two independent units tangents $\boldsymbol{t}_1$ and $\boldsymbol{t}_2$ we can construct equations for the two tangential components of $\boldsymbol{H}$ along these two directions: $H_{t_1}$ and $H_{t_2}$. See (\ref{eq:HG_5N_by_5N}) below for details.

We see that our formulation for PEC problems for $\boldsymbol{H}$ is slightly more complex than our formulation for $\boldsymbol{E}$ because of the need to use the boundary condition for $\boldsymbol{E}$ to find boundary conditions for $\boldsymbol{H}$ in (\ref{eq:Et_EQ_0}). 

So for solving practical PEC problems, (\ref{eq:EH_wave_eqn_a}) and (\ref{eq:div_EH_a}) should be used to solve for $\boldsymbol{E}$, and $\boldsymbol{H}$ can be found subsequently from $\boldsymbol{E}$ \emph{via} Maxwell's equations. However, it is also possible to solve directly for $\boldsymbol{H}$ using (\ref{eq:EH_wave_eqn_b}), (\ref{eq:div_EH_b}), (\ref{eq:Hn_eq_0}) and (\ref{eq:Et_EQ_0}). We will provide illustrations of these points in the results section.

\subsection{\label{sec:PEC_results} PEC results}

The scalar Helmholtz equations (\ref{eq:EH_wave_eqn_a}) and (\ref{eq:div_EH_a}) for the three Cartesian  components of $\boldsymbol{E}$ and the scalar function ($\boldsymbol{r} \cdot \boldsymbol{E}$) can be formulated as a system of linear equations that is the discretized representation of four non-singular boundary integral equations of the Helmholtz equations using (\ref{eq:NS_BIE}). The total field, $\boldsymbol{E}$, can be written as the sum of the incident and scattered fields: $\boldsymbol{E} = \boldsymbol{E}^{inc} + \boldsymbol{E}^{scat}$. Since the known incident field, $\boldsymbol{E}^{inc}$, such as a plane wave, satisfies (\ref{eq:EH_wave_eqn_a}) and (\ref{eq:div_EH_a}), we can solve for the unknown scattered field, $\boldsymbol{E}^{scat}$ that satisfies the Sommerfeld radiation condition at infinity.

On the surface of a PEC object, we work in terms of the normal and tangential components of the scattered field: $\boldsymbol{E}^{scat}=\boldsymbol{E}^{scat}_{n}+\boldsymbol{E}^{scat}_{t}$. Since the tangential component of the total field, $\boldsymbol{E}$ must vanish on the surface of a PEC, then the tangential components of the scattered and incident fields must cancel. Thus the Cartesian components of the scattered field, $\boldsymbol{E}^{scat}$ on the surface of a PEC can be expressed in terms of the known tangential components of the incident field, $\boldsymbol{E}^{inc}_t = (E^{inc}_{t,x}, E^{inc}_{t,y}, E^{inc}_{t,z})$, the components of the surface normal, $\boldsymbol{n} = (n_x, n_y, n_z)$ and the unknown normal component of the scattered field, $\boldsymbol{E}^{scat}_{n} = E^{scat}_n \boldsymbol{n}$ as follows:
\begin{eqnarray} \label{eq:Escat_En}
 E_{\alpha}^{scat} &=& E_{n}^{scat} \; n_{\alpha} + E_{t,{\alpha}}^{scat} \nonumber \\  
    &=& E_{n}^{scat} \; n_{\alpha} - E_{t,{\alpha}}^{inc}, \qquad (\alpha = x, y, z).  
\end{eqnarray}

We discretize the surface, $S$ using quadratic triangular area elements where each element is bounded by 3 nodes on the vertices and 3 nodes on the edges for a total of $N$ nodes on the surface. The coordinates of a point within each element and the function values at that point are obtained by quadratic interpolation from the values at the nodes on the element~\cite{Klaseboer_2017, Sun_2016}.

The solution of  (\ref{eq:EH_wave_eqn_a}) and (\ref{eq:div_EH_a}) for components of the scattered field, $\boldsymbol{E}^{scat}$ and $(\boldsymbol{r} \cdot \boldsymbol{E}^{scat})$ on the surface are expressed in terms of the field values at the $N$ surface nodes. The surface integrals in (\ref{eq:NS_BIE}) can be expressed as a system of linear equations in which the elements of the matrices $\cal{H}$ and $\cal{G}$ are the results of integrals over the surface elements involving the unknown $4N$-vector ($E^{scat}_x, E^{scat}_y, E^{scat}_z, \boldsymbol{r}\cdot\boldsymbol{E}^{scat}$). The integral over each surface element can be calculated accurately and efficiently using standard Gauss quadrature since the integrands have no singularities. The resulting linear system can be written as
\begin{subequations}  \label{eq:HG_matrix_eq}
\begin{eqnarray}
 {\cal{H}} \cdot E^{scat}_{\alpha} &=& {\cal{G}} \cdot (\partial{E^{scat}_\alpha}/\partial{n}), \; (\alpha = x, y, z) \quad  \label{eq:HG_matrix} \\
  {\cal{H}} \cdot (\boldsymbol{r}\cdot \boldsymbol{E}^{scat}) &=& {\cal{G}} \cdot [\partial{(\boldsymbol{r}\cdot \boldsymbol{E}^{scat})}/\partial{n}]. \;\label{eq:HG_matrix_rdotE}
\end{eqnarray}
\end{subequations}
For the left hand sides of (\ref{eq:HG_matrix}), we use (\ref{eq:Escat_En}) to eliminate the Cartesian components: $E^{scat}_{\alpha}$, ($\alpha = x, y, z$) in favor of the normal component, ($E^{scat}_{n} \boldsymbol{n}$) and the tangential component of the known incident field, $\boldsymbol{E}_{t}^{inc}$. For (\ref{eq:HG_matrix_rdotE}), we use (\ref{eq:Escat_En}) to write
\begin{subequations} \label{eq:r_dot_Escat}
\begin{eqnarray}
  \boldsymbol{r}\cdot \boldsymbol{E}^{scat} &=&  (\boldsymbol{r}\cdot \boldsymbol{n})E^{scat}_{n} + (\boldsymbol{r}\cdot \boldsymbol{E}_{t}^{scat}) \nonumber \\
    &=& (\boldsymbol{r}\cdot \boldsymbol{n})E^{scat}_{n}-(\boldsymbol{r}\cdot \boldsymbol{E}_{t}^{inc})
\\
  \frac{\partial{(\boldsymbol{r}\cdot \boldsymbol{E}^{scat})}}{\partial{n}} &=& E^{scat}_{n}+\boldsymbol{r}\cdot \frac{\partial{\boldsymbol{E}^{scat}}}{\partial{n}}.
\end{eqnarray}
\end{subequations}
Thus using (\ref{eq:Escat_En}) and (\ref{eq:r_dot_Escat}), (\ref{eq:HG_matrix})  can be expressed in terms of the 3 components of the normal derivative $\partial{\boldsymbol{E}^{scat}}/\partial{n}$ of the scattered field and the normal component of the scattered field, $E^{scat}_{n}$. Applying these results at the $N$ nodes of the surface, we obtain a $4N\times 4N$ system of linear equations for the 4$N$-vector: $(\partial{E^{scat}_x}/\partial{n}, \partial{E^{scat}_y}/\partial{n}, \partial{E^{scat}_z}/\partial{n}, E^{scat}_n )$ of unknowns on the surface to be determined
\begin{align} \label{eq:HG_4N_by_4N}
 \begin{bmatrix}
  \hspace{-0.4 cm}  -{\cal{G}} & 0 & 0 & {\cal{H}}n_x \\
    0 & -{\cal{G}} & 0 & {\cal{H}}n_y \\
    0 & 0 & -{\cal{G}} & {\cal{H}}n_z \\ 
    -{\cal{G}}x & -{\cal{G}}y & -{\cal{G}}z & {\cal{Y}} \end{bmatrix}
    \hspace{-0.2  cm}
  \left[ \begin{array}{c} \partial_n E^{scat}_x \\ \partial_n E^{scat}_y \\ \partial_n E^{scat}_z \\ E^{scat}_n \end{array} \right] 
\hspace{-0.1 cm} = \hspace{-0.1 cm} \left[ \begin{array}{c}  {\cal{H}}  E_{t,x}^{inc}\\ {\cal{H}}  E_{t,y}^{inc} \\ {\cal{H}}  E_{t,z}^{inc} \\ {\cal{Z}}^E 
 \end{array} \right]
\end{align}
where ${\cal{Y}} \equiv -{\cal{G}}+{\cal{H}}(\boldsymbol{r}\cdot \boldsymbol{n})$, ${\cal{Z}}^E \equiv {\cal{H}} (\boldsymbol{r}\cdot \boldsymbol{E}_{t}^{inc})$ and $ \partial_n E^{scat}_\alpha \equiv \boldsymbol{n} \cdot \nabla E^{scat}_\alpha$, ($\alpha = x,y,z$). This linear system gives the scattered field, $\boldsymbol{E}_{scat}$ from the PEC surface in terms of the incident field, $\boldsymbol{E}_{inc}$.

The linear system for the $\boldsymbol{H}$ field can be obtained from (\ref{eq:EH_wave_eqn_b}), (\ref{eq:Hn_eq_0}) and (\ref{eq:Ep_to_H}). However, unlike the linear system for $\boldsymbol{E}$, the tangential boundary condition (\ref{eq:Ep_to_H}) for $\boldsymbol{H}$ on the surface gives rise to two unknowns, being the components of the $\boldsymbol{H}$ along the directions of two orthogonal unit tangents $\boldsymbol{t}_1$ and $\boldsymbol{t}_2$ that are related to the unit normal $\boldsymbol{n}$ by $\boldsymbol{t}_2 \equiv \boldsymbol{t}_1 \times \boldsymbol{n}$. In this case, there are $5N$ unknowns comprising the $2N$ unknowns for the tangential components of the scattered field $\boldsymbol{H}^{scat}$ and $3N$ unknowns for the components of ($\partial \boldsymbol{H}^{scat}/ \partial n$) to be determined by the following $5N \times 5N$ linear system
\begin{align} \label{eq:HG_5N_by_5N}
 \begin{bmatrix}
   \hspace{-0.2 cm} -{\cal{G}} & 0 & 0 & {\cal{H}}t_{1x} & {\cal{H}}t_{2x} \\
    0 & -{\cal{G}} & 0 & {\cal{H}}t_{1y} & {\cal{H}}t_{2y} \\
    0 & 0 & -{\cal{G}} & {\cal{H}}t_{1z} & {\cal{H}}t_{2z} \\ 
    -t_{1x} & -t_{1y} & -t_{1z} & - {\cal{K}}_1 & 0 \\
    -t_{2x} & -t_{2y} & -t_{2z} & 0 & -{\cal{K}}_2
  \end{bmatrix}
  \hspace{-0.2 cm}
  \left[ \begin{array}{c} \partial_n H^{scat}_x \\ \partial_n H^{scat}_y \\ \partial_n H^{scat}_z \\ H^{scat}_{t_1} \\  H^{scat}_{t_2} \end{array} \right]  \hspace{-0.1 cm} = \hspace{-0.1 cm}
   \left[ \begin{array}{c}  {\cal{H}}  H_{n,x}^{inc}\\
    {\cal{H}}  H_{n,y}^{inc} \\ 
   {\cal{H}}  H_{n,z}^{inc} \\ 
    {\cal{Z}}^H_1\\ 
   {\cal{Z}}^H_2 \end{array} \right]
\end{align}
where ${\cal{Z}}^H_j \equiv \boldsymbol{t}_{j} \cdot  \partial_n  \boldsymbol{H}^{inc} + \kappa_j \boldsymbol{n} \cdot \boldsymbol{H}^{inc}$, with $\kappa_j$ being the local curvature of the surface, $S$ along the tangential direction $\boldsymbol{t}_j$ with $j = 1, 2$. As in (\ref{eq:HG_4N_by_4N}), we use the notation: $\partial_n (\cdots) \equiv \boldsymbol{n} \cdot \nabla (\cdots)$. The sub-matrix ${\cal{K}}_j$ is diagonal with the non-zero diagonal entries populated by the local curvature $\kappa_j$. The entries involving the local curvature $\kappa_j$ and ${\cal{Z}}^H_j$ originate from the $\boldsymbol{n} \cdot (\boldsymbol{t} \cdot \nabla) \boldsymbol{H}$ term in (\ref{eq:Ep_to_H}) that involves tangential derivatives of $\boldsymbol{H}$. They can be derived with some algebraic manipulation using  the following relations from differential geometry between the surface normal, $\boldsymbol{n}$, surface tangent, $\boldsymbol{t}_j$ and local curvature, $\kappa_j$: $\partial \boldsymbol{n} / \partial \boldsymbol{t}_j = \kappa_j \boldsymbol{t}_j$, $\partial \boldsymbol{t}_j / \partial \boldsymbol{t}_j = -\kappa_j \boldsymbol{n}$, $\partial \boldsymbol{t}_1 / \partial \boldsymbol{t}_2 = \boldsymbol{0}$ and $\partial \boldsymbol{t}_2 / \partial \boldsymbol{t}_1 = \boldsymbol{0}$. This linear system (\ref{eq:HG_5N_by_5N}) then gives the scattered magnetic field, $\boldsymbol{H}^{scat}$ in terms of the incident field, $\boldsymbol{H}^{inc}$ at a PEC surface.

Apart from obtaining the surface fields on a PEC scatterer by solving (\ref{eq:HG_4N_by_4N}) or (\ref{eq:HG_5N_by_5N}), we also use the non-singular boundary integral formalism to calculate field values anywhere in the 3D domain and field gradients at the surface~\cite{Sun_2015b} that are often sought in quantifying surface field enhancement effects in micro-photonics applications. As mentioned earlier, we discretize the surface, $S$ using quadratic triangular area elements where each element is bounded by 3 nodes on the vertices and 3 nodes on the edges for a total of $N$ nodes on the surface. In evaluating the surface integrals over each element, the coordinates of a point within each element and the function values at that point are obtained by quadratic interpolation from the values at the nodes on the element that are the unknowns to be solved.

To facilitate discussion of sample results to follow, we denote our two methods of solving scattering problems by PEC objects as:
\begin{description}
\item[PEC-E] ~being based on (\ref{eq:EH_wave_eqn_a}), (\ref{eq:div_EH_a}) and $\boldsymbol{E}_{t} = \boldsymbol{0}$ on $S$, that gives the $4N\times 4N$ linear system (\ref{eq:HG_4N_by_4N}) to solve for the $\boldsymbol{E}$ field, and
\item[PEC-H] ~being based on (\ref{eq:EH_wave_eqn_b}), (\ref{eq:div_EH_b}), (\ref{eq:Hn_eq_0}) and (\ref{eq:Ep_to_H}), that gives the $5N\times 5N$ linear system (\ref{eq:HG_5N_by_5N}) to solve for the $\boldsymbol{H}$ field.
\end{description}

In the next two subsections we first benchmark our approach against the analytic Mie solution of the scattering of a plane wave by a PEC sphere and then we present results for scattering by more complex objects such as multiple PEC spheres and a high geometric aspect ratio PEC scatterer, with particular focus on the behavior of $\boldsymbol{E}$ and $\boldsymbol{H}$ on or near the surface of the scatterers. In all examples, the incident field is plane polarized in the $x$-direction: $\boldsymbol{E}^{inc} = (1,0,0) \exp(ikz)$ and it travels in the positive $z$-direction with $\boldsymbol{k} = (0,0,k)$.  
\begin{figure} [ph]
  \centering{}
  \subfloat[]{ \includegraphics[width=2.6in]{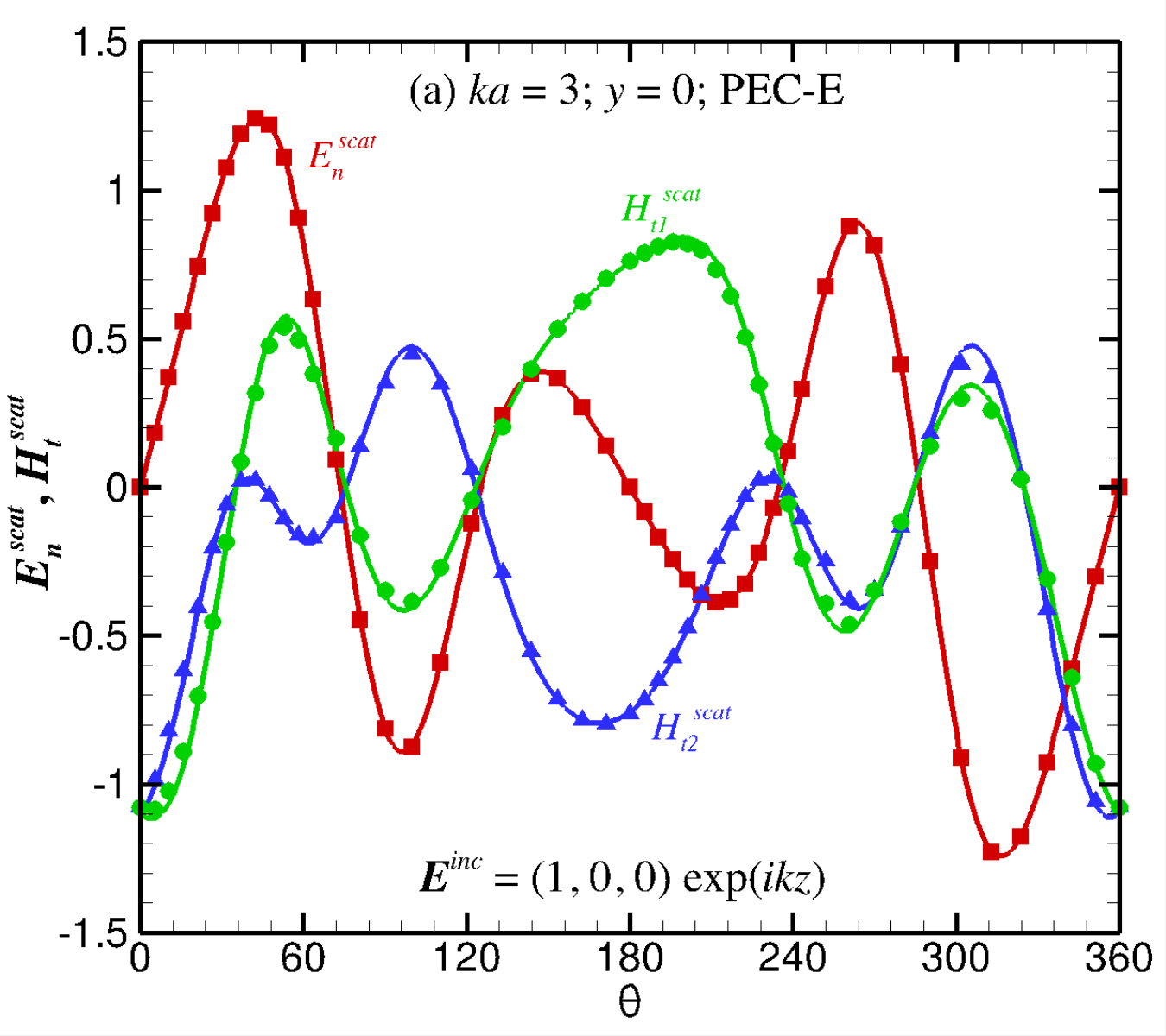} }\\
  \subfloat[]{ \includegraphics[width=2.6in]{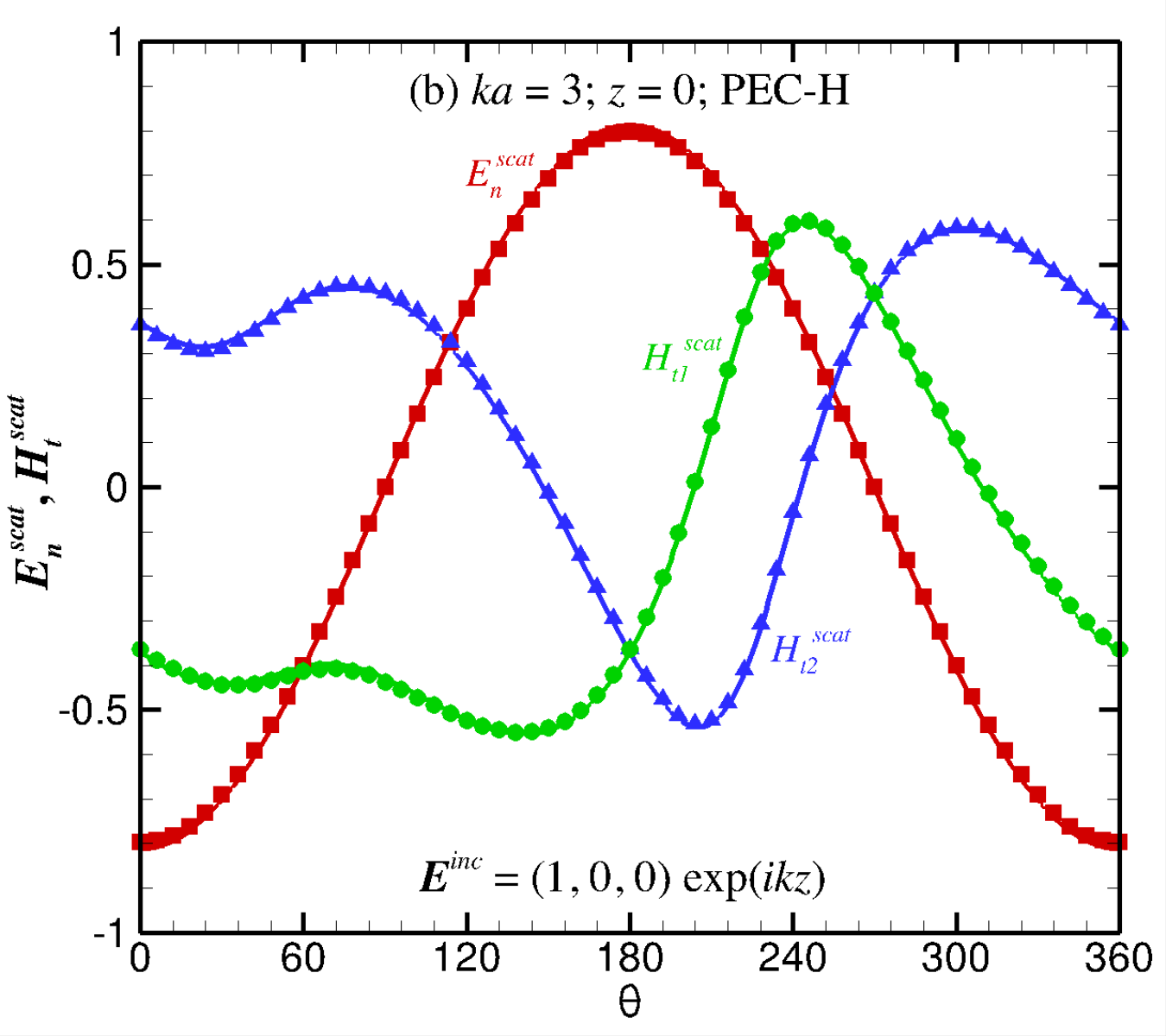} }\\
  \subfloat[]{ \includegraphics[width=2.6in]{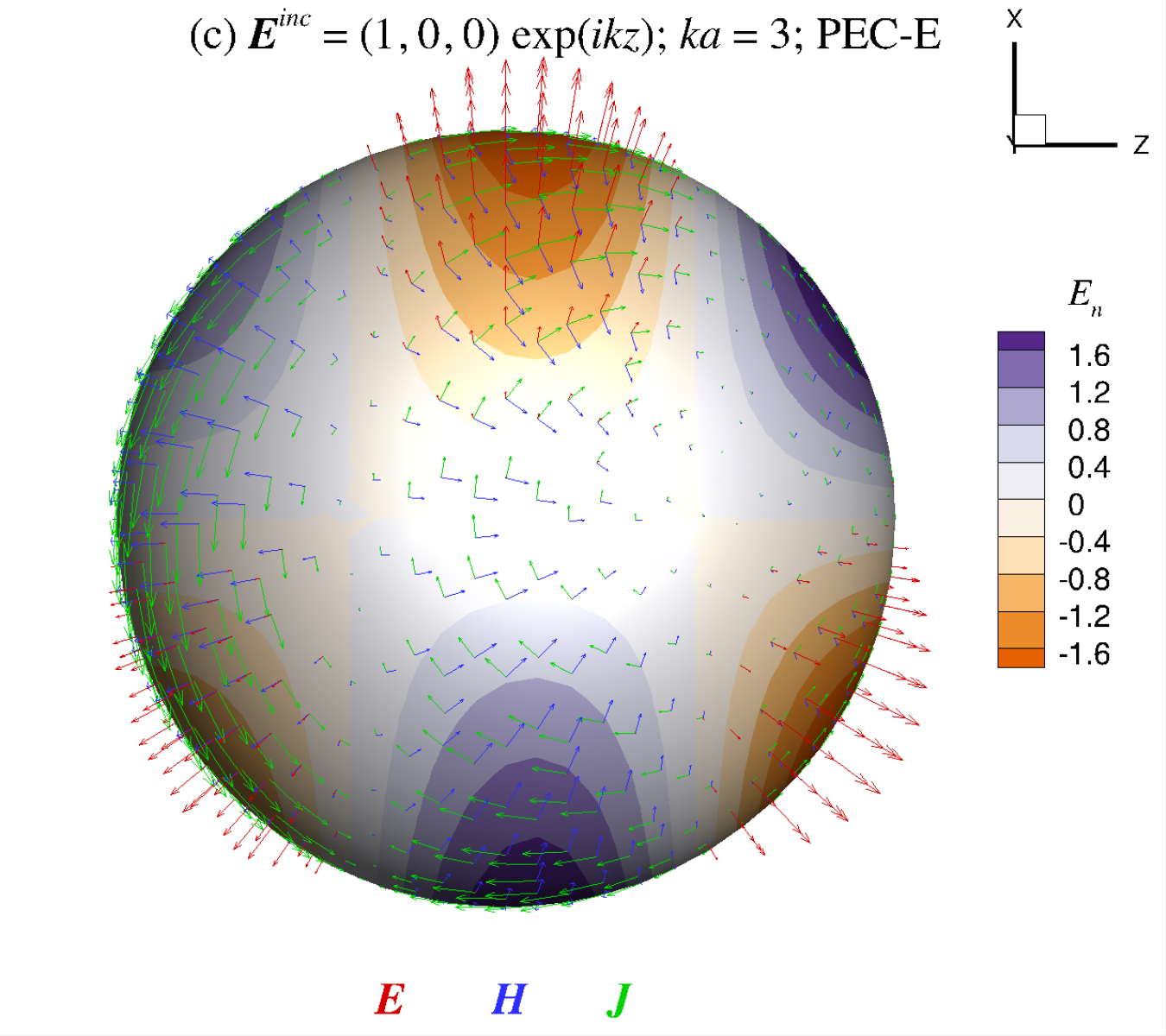} } 
  \caption{\label{Fig:PEC_Mie_EH} (Color on-line) Scattering of a plane wave by a perfect electrical conducting (PEC) sphere: the normal component of the scattered electric field, $E_n^{scat}$ and the tangential components of the scattered magnetic field, $H_{t_1}^{scat}$ and $H_{t_2}^{scat}$ on the sphere at (a) $y=0$ and (b) $z=0$ obtained from the PEC-E and PEC-H methods of the field-only formulation (symbols) and Mie theory (lines). (c) The total surface fields $\boldsymbol{E}$ and $\boldsymbol{H}$ and the induced electric current density, $\boldsymbol{J}$ obtained using the PEC-E approach.  The magnitude of the normal component of the total field, $E_n$ is given on a color scale. (Animation on-line)} 
\end{figure}
\subsubsection{\label{sec:PEC_Mie_sphere} Perfect Electrical Conducting (PEC) Sphere vs Mie}

In Fig.~\ref{Fig:PEC_Mie_EH}a and~\ref{Fig:PEC_Mie_EH}b we show variations of components of the scattered electric field, $\boldsymbol{E}^{scat}$ and magnetic field, $\boldsymbol{H}^{scat}$ along meridians in the planes $y=0$ and $z=0$ on the surface of a PEC sphere of radius, $a$ resulting from an incident plane wave. We compare our present field-only formulation using the PEC-E and PEC-H approaches with the analytic Mie theory~\cite{vandeHulst_1957,Liou_1977} at $ka = 3$. The tangential components of $\boldsymbol{H}^{scat}$ are given along the unit vectors: $\boldsymbol{t}_2 = (n_y-n_z, n_z-n_x, n_x-n_y)$ and $\boldsymbol{t}_1 = \boldsymbol{n} \times \boldsymbol{t}_2$. For an incident field $\boldsymbol{E}^{inc} = (1,0,0) \exp(ikz)$ the absolute difference between the two methods is less than 0.01 with 720 quadratic elements and $N=1442$ nodes .
 \begin{figure} [t]
  \centering{}
  \subfloat[]{ \includegraphics[width=3in]{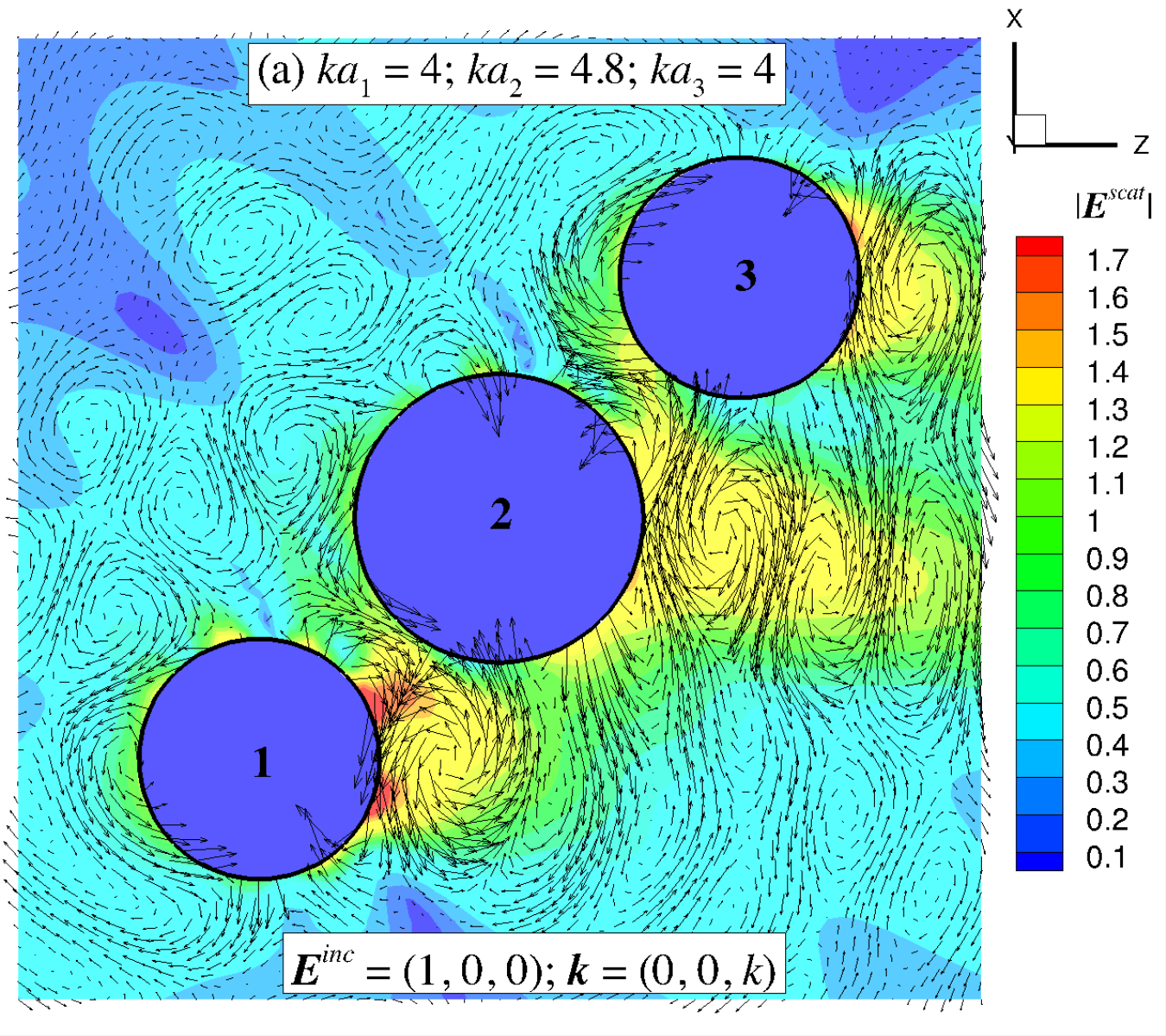} }\\
  \subfloat[]{ \includegraphics[width=3in]{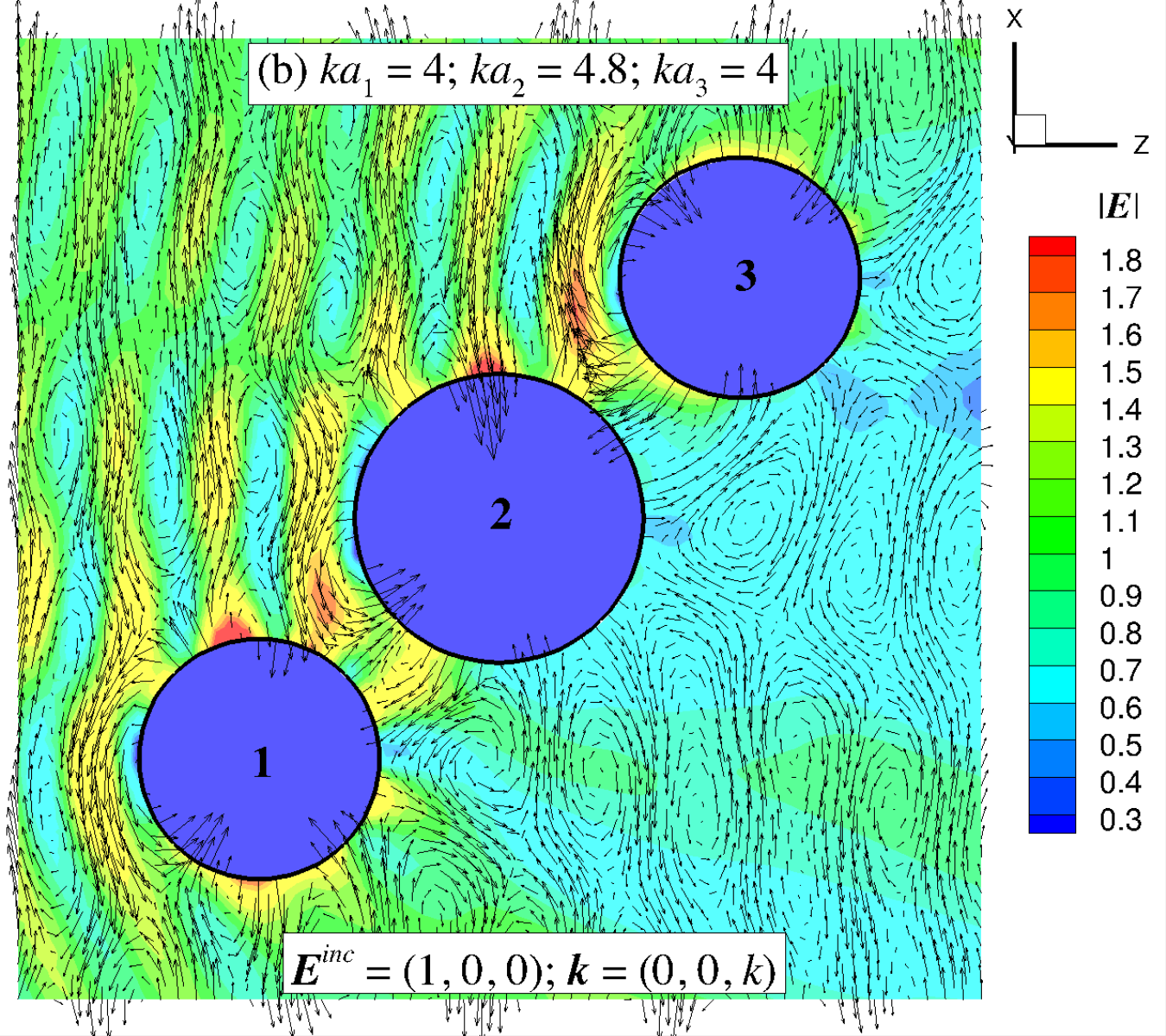} } 
\caption{\label{Fig:PEC3spheres}  (Color on-line) (a) The scattered electric field, $\boldsymbol{E}^{scat}$ and (b) the total field, $\boldsymbol{E}$ in the plane $y=0$ around 3 perfect conducting spheres due to an incident electric field, $\boldsymbol{E}^{inc} = (1,0,0) \exp (ikz)$ with $ka_{1} = ka_{3}= 4$ and $ka_{2}=4.8$.  Results are obtained using 1442 nodes and 720 quadratic elements on each sphere. (Animation on-line)} 
\end{figure}

In Fig.~\ref{Fig:PEC_Mie_EH}c, we show vector plots of the total fields $\boldsymbol{E}$, $\boldsymbol{H}$ and the induced surface current density, $\boldsymbol{J} = \boldsymbol{H} \times \boldsymbol{n}$ on the surface of the same PEC sphere. The induced surface charge density that is proportional to the normal component of the total field, $E_n$ at the surface is shown on the color scale.

\subsubsection{\label{sec:Complex_PEC_Objects} Near field around complex PEC objects}

In Fig.~\ref{Fig:PEC3spheres} we show the field in the plane $y=0$ around three co-linear PEC spheres that are oriented at $45^{\circ}$ to the propagation direction of the incident plane wave. The centers of the spheres are located at $(-2a, 0, -2a)$, $(0, 0, 0)$ and $(2a, 0, 2a)$. These fields are obtained by post-processing the surface field values obtained from solving the PEC-E equations. Interference between the scattered field shown in Fig.~\ref{Fig:PEC3spheres}a and the incident field gives the complex total field structure on the downstream side of the scatterers in Fig.~\ref{Fig:PEC3spheres}b. The magnitudes of the scattered and total field strength, illustrated on a colour scale, quantify the shielding effect of this 3-sphere structure. 

\begin{figure}[t]
\centering{}\includegraphics[width=3in]{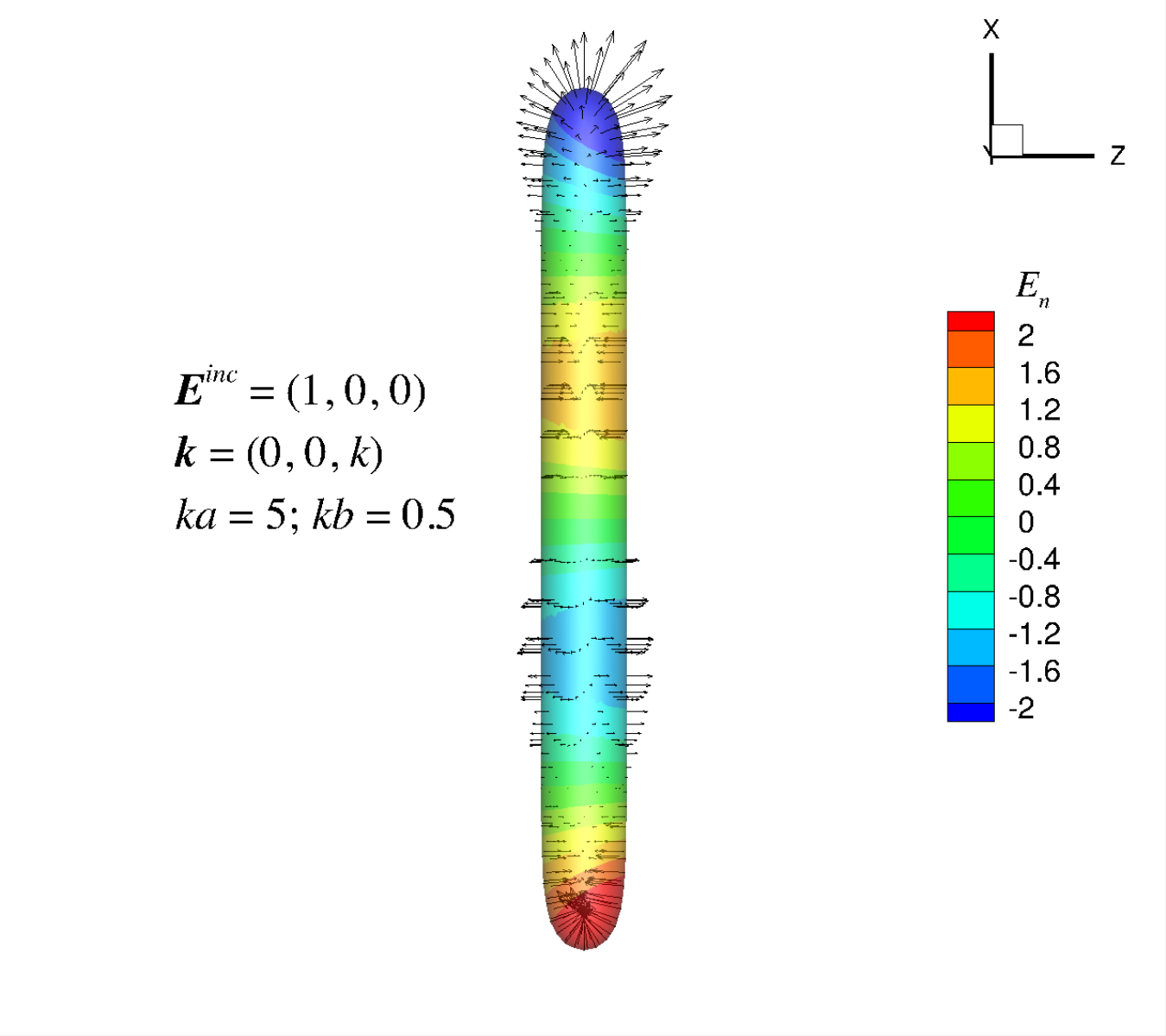} \caption{\label{Fig:PECneedle} (Color on-line) The total field on the surface of a perfect conducting needle with length, $2a$ and width $2b$ due to an incident electric field, $\boldsymbol{E}^{inc} = (1,0,0) \exp (ikz)$ with $ka = 5$ and $kb=0.5$. The magnitude of the field, $E_n$ is given by the color scale. The results are obtained using the present field-only formulation with $N=1002$ nodes and 500 quadratic elements. An analytic equation~\cite{Chwang_1974} is used for the surface of the needle. (Animation on-line)}
\end{figure}

To illustrate the ability of our field-only formulation to handle scatterers that have widely varying characteristic dimensions, we show the surface field on a PEC needle that has a large length, $2a$ to width, $2b$ ratio in Fig.~\ref{Fig:PECneedle} at $ka = 5$ and $kb = 0.5$.

Since we have assumed harmonic time dependence, animations that illustrate the phase behavior of results in Fig.~\ref{Fig:PEC_Mie_EH}c, Fig.~\ref{Fig:PEC3spheres} and Fig.~\ref{Fig:PECneedle} are given in the supplementary material. 

\section{\label{sec:DIEL} Dielectric scatterers}

\subsection{\label{sec:DIEL_formulation} Dielectric formulation}

For scattering by a dielectric object we denote the domain containing the \emph{incident} fields: $\boldsymbol{E}^{inc}$ and $\boldsymbol{H}^{inc}$ and \emph{scattered} fields: $\boldsymbol{E}^{scat}$ and $\boldsymbol{H}^{scat}$ as the \emph{outside} with material constants $\epsilon_{out}$ and $\mu_{out}$ and corresponding wave number $k_{out}$ in the Helmholtz equations. A boundary surface, $S$ separates this from the \emph{inside} with the  \emph{transmitted} fields: $\boldsymbol{E}^{tran}$ and $\boldsymbol{H}^{tran}$ with material constants $\epsilon_{in}$ and $\mu_{in}$, and wave number $k_{in}$.

For a given incident field, we solve (\ref{eq:EH_wave_eqn}) and (\ref{eq:div_EH}) for wave numbers corresponding to the \emph{outside} and \emph{inside} domains (assuming they are both source free) and use the continuity of tangential components of $\boldsymbol{E}$ and $\boldsymbol{H}$ at the boundary, $S$. The continuity of the normal components of $\boldsymbol{D}$ and $\boldsymbol{B}$ follows from Maxwell's equations.~\cite{StrattonChu_1939} 

Since in our formulation both $\boldsymbol{E}$ and $\boldsymbol{H}$ satisfy the same equations (\ref{eq:EH_wave_eqn}) and (\ref{eq:div_EH}) and similar boundary conditions, it is only necessary to formulate the solution for $\boldsymbol{E}$ as the solution for $\boldsymbol{H}$ can be found by replacing $\boldsymbol{H}$ with $\boldsymbol{E}$ and interchanging $\epsilon$ and $\mu$.
  
The boundary integral equations for the electric field will involve 9 unknown functions on the surface, 6 of which are normal derivatives of the Cartesian components of the scattered and transmitted fields: $\partial_n \boldsymbol{E}^{scat} \equiv \boldsymbol{n} \cdot \nabla \boldsymbol{E}^{scat}$ and $\partial_n \boldsymbol{E}^{tran} \equiv \boldsymbol{n} \cdot \nabla \boldsymbol{E}^{tran}$. For the remaining 3 unknowns, we can either choose to solve for the 3 components of the scattered field ($E^{scat}_n, E^{scat}_{t_1}, E^{scat}_{t_2}$) along the normal, $\boldsymbol{n}$ and tangential, $\boldsymbol{t}_1$ and $\boldsymbol{t}_2$, directions for which the linear system is given in Table~\ref{Tbl:ls}a, or choose to solve for the 3 components the transmitted field ($E^{scat}_n, E^{scat}_{t_1}, E^{scat}_{t_2}$), for which the linear system is given in Table~\ref{Tbl:ls}b.


The surface integral matrices ${\cal{G}}$ and ${\cal{H}}$ in Table~\ref{Tbl:ls}a and Table~\ref{Tbl:ls}b are calculated using the Green's function, $G(\boldsymbol{r},\boldsymbol{r}_0)$ in (\ref{eq:Greens_function}) with $k = k_{out}$ and the surface integral matrics ${\cal{G}}_{in}$ and ${\cal{H}}_{in}$ are calculated using $G(\boldsymbol{r},\boldsymbol{r}_0)$ with $k = k_{in}$.

The corresponding equations for the scattered and transmitted $\boldsymbol{H}$ fields can obtained from Table~\ref{Tbl:ls}a and \ref{Tbl:ls}b by replacing the components of $\boldsymbol{E}$ by the corresponding components of $\boldsymbol{H}$ and interchanging $\epsilon$ and $\mu$. For example, in Table~\ref{Tbl:ls}c, we give the linear system that determines $\partial \boldsymbol{H}^{scat}/\partial n$,  $\partial \boldsymbol{H}^{tran}/\partial n$ and the \emph{scattered} field, $\boldsymbol{H}^{scat}$. These results, together with (\ref{eq:HG_4N_by_4N}) and (\ref{eq:HG_5N_by_5N}) for perfect electrical conducting scatterers, are applicable for any form of incident field that appear in the vectors on the right hand side. 

The matrix equations are presented in a way that best reflect the algebraic structure of our field-only formulation in terms of Helmholtz equations in (\ref{eq:EH_wave_eqn}) and (\ref{eq:div_EH}). The first three lines of the linear system in Table~\ref{Tbl:ls}a are, respectively, the discretized version of Helmholtz equation for the $x$-, $y$- and $z$-components of $\boldsymbol{E}^{scat}$. The 5th to 7th lines are, respectively, the discretized version of Helmholtz equation for the $x$-, $y$- and $z$-components of $\boldsymbol{E}^{tran}$ after using the continuity of the tangential components of $\boldsymbol{E}$ and the normal component of $\boldsymbol{D}$. The 4th line originates from $\nabla \cdot\boldsymbol{E}^{scat} = 0$, see (\ref{eq:div_EH}), and finally lines 8 and 9 express the continuity of the tangential components of $\boldsymbol{H}$ across the boundary in terms of $\boldsymbol{E}$.
\begin{table*}[htp]\centering
    \captionsetup{justification=justified, singlelinecheck=false}
    \caption{The linear systems}\label{Tbl:ls}
    \ra{1.3}
    \begin{ruledtabular}
    \begin{tabular}{l}
        \toprule
        (a) The linear system for $\partial \boldsymbol{E}^{scat}/\partial n$,  $\partial \boldsymbol{E}^{tran}/\partial n$ and the \emph{scattered} field, $\boldsymbol{E}^{scat}$:  \\
        \noalign{\smallskip}\noalign{\smallskip}
      $\begin{bmatrix}
  -{\cal{G}} & 0 & 0  & {\cal{H}}n_x & 0 & 0 & 0 & {\cal{H}}t_{1x} & {\cal{H}}t_{2x}  \\
    0 & -{\cal{G}} & 0 & {\cal{H}}n_y & 0 & 0 & 0 & {\cal{H}}t_{1y} & {\cal{H}}t_{2y}  \\
    0 & 0 & -{\cal{G}} & {\cal{H}}n_z & 0 & 0 & 0 & {\cal{H}}t_{1z} & {\cal{H}}t_{2z}   \\ 
   -{\cal{G}} x & -{\cal{G}} y & -{\cal{G}} z & -{\cal{G}}+{\cal{H}}(\boldsymbol{r}\cdot \boldsymbol{n}) & 0 & 0  &  0 & {\cal{H}}(\boldsymbol{r}\cdot \boldsymbol{t}_{1})  & {\cal{H}}(\boldsymbol{r}\cdot \boldsymbol{t}_{2}) \\
    0 & 0 & 0 & \epsilon_{oi} {\cal{H}}_{in}n_x & -{\cal{G}}_{in} & 0 & 0 & {\cal{H}}_{in}t_{1x} & {\cal{H}}_{in}t_{2x}  \\
    0 & 0 & 0 & \epsilon_{oi} {\cal{H}}_{in}n_y & 0 & -{\cal{G}}_{in} & 0 & {\cal{H}}_{in}t_{1y} & {\cal{H}}_{in}t_{2y}  \\ 
    0 & 0 & 0 & \epsilon_{oi} {\cal{H}}_{in}n_z & 0 & 0 & -{\cal{G}}_{in} & {\cal{H}}_{in}t_{1z} & {\cal{H}}_{in}t_{2z}  \\
   - \mu_{io} t_{1x} & - \mu_{io} t_{1y} & - \mu_{io} t_{1z} & (\mu_{io} - \epsilon_{oi}) \partial_{t_1} & t_{1x} & t_{1y} & t_{1z} & (1 - \mu_{io}) {{\cal{K}}_1} & 0     \\
    -\mu_{io} t_{2x} & - \mu_{io} t_{2y} & - \mu_{io} t_{2z} & (\mu_{io} - \epsilon_{oi}) \partial_{t_2} & t_{2x} & t_{2y} & t_{2z} & 0 & (1 - \mu_{io}) {{\cal{K}}_2}
  \end{bmatrix} \hspace{-0.2 cm}
  \left[ \begin{array}{c} \partial_n E^{scat}_x \\ \partial_n E^{scat}_y \\ \partial_n E^{scat}_z \\ E^{scat}_n \\ \partial_n E^{tran}_x \\ \partial_n E^{tran}_y \\ \partial_n E^{tran}_z \\ E^{scat}_{t_1} \\ E^{scat}_{t_2}  \end{array} \right] 
   \hspace{-0.1 cm} =  \hspace{-0.1 cm} 
  \left[ \begin{array}{c}   0 \\ 0 \\ 0 \\ 0 \\  {\cal{A}}_x  \\ {\cal{A}}_y  \\  {\cal{A}}_z  \\ {\cal{B}}_1 \\ {\cal{B}}_2  
 \end{array} \right] $\\
 \\
      (b) The linear system for $\partial \boldsymbol{E}^{scat}/\partial n$,  $\partial \boldsymbol{E}^{tran}/\partial n$ and the \emph{transmitted} field, $\boldsymbol{E}^{tran}$:  \\
         \noalign{\smallskip}\noalign{\smallskip}
        $\begin{bmatrix}
   -{\cal{G}} & 0 & 0  & \epsilon_{io} {\cal{H}}n_x & 0 & 0 & 0 & {\cal{H}}t_{1x} & {\cal{H}}t_{2x}  \\
    0 & -{\cal{G}} & 0 & \epsilon_{io} {\cal{H}}n_y & 0 & 0 & 0 & {\cal{H}}t_{1y} & {\cal{H}}t_{2y}  \\
    0 & 0 & -{\cal{G}} & \epsilon_{io} {\cal{H}}n_z & 0 & 0 & 0 & {\cal{H}}t_{1z} & {\cal{H}}t_{2z}   \\ 
   0 & 0 & 0 & {\cal{H}}_{in}(\boldsymbol{r}\cdot \boldsymbol{n}) -{\cal{G}}_{in} & -{\cal{G}}_{in} x & -{\cal{G}}_{in} y & -{\cal{G}}_{in} z & {\cal{H}}_{in} (\boldsymbol{r}\cdot \boldsymbol{t}_{1})  & {\cal{H}}_{in} (\boldsymbol{r}\cdot \boldsymbol{t}_{2}) \\
    0 & 0 & 0 & {\cal{H}}_{in}n_x & -{\cal{G}}_{in} & 0 & 0 & {\cal{H}}_{in}t_{1x} & {\cal{H}}_{in}t_{2x}  \\
    0 & 0 & 0 & {\cal{H}}_{in}n_y & 0 & -{\cal{G}}_{in} & 0 & {\cal{H}}_{in}t_{1y} & {\cal{H}}_{in}t_{2y}  \\ 
    0 & 0 & 0 & {\cal{H}}_{in}n_z & 0 & 0 & -{\cal{G}}_{in} & {\cal{H}}_{in}t_{1z} & {\cal{H}}_{in}t_{2z}  \\
   - \mu_{io} t_{1x} & - \mu_{io} t_{1y} & - \mu_{io} t_{1z} & ( \epsilon_{io} \mu_{io} - 1) \partial_{t_1} & t_{1x} & t_{1y} & t_{1z} & (1 - \mu_{io}) {{\cal{K}}_1} & 0     \\
    -\mu_{io} t_{2x} & - \mu_{io} t_{2y} & - \mu_{io} t_{2z} & ( \epsilon_{io} \mu_{io} - 1) \partial_{t_2} & t_{2x} & t_{2y} & t_{2z} & 0 & (1 - \mu_{io}) {{\cal{K}}_2}
  \end{bmatrix}  \hspace{-0.2 cm}
  \left[ \begin{array}{c} \partial_n E^{scat}_x \\ \partial_n E^{scat}_y \\ \partial_n E^{scat}_z \\ E^{tran}_n \\ \partial_n E^{tran}_x \\ \partial_n E^{tran}_y \\ \partial_n E^{tran}_z \\ E^{tran}_{t_1} \\ E^{tran}_{t_2}  \end{array} \right] 
   \hspace{-0.1 cm} =  \hspace{-0.1 cm}
  \left[ \begin{array}{c}   {\cal{C}}_x \\ {\cal{C}}_y \\ {\cal{C}}_z \\ 0 \\  0  \\ 0  \\  0  \\ {\cal{D}}_1 \\ {\cal{D}}_2  
 \end{array} \right]$ \\
 \\
        (c) The linear system for $\partial \boldsymbol{H}^{scat}/\partial n$,  $\partial \boldsymbol{H}^{tran}/\partial n$ and the \emph{scattered} field, $\boldsymbol{H}^{scat}$:   \\
         \noalign{\smallskip}\noalign{\smallskip}
        $\begin{bmatrix}
  -{\cal{G}} & 0 & 0  & {\cal{H}}n_x & 0 & 0 & 0 & {\cal{H}}t_{1x} & {\cal{H}}t_{2x}  \\
    0 & -{\cal{G}} & 0 & {\cal{H}}n_y & 0 & 0 & 0 & {\cal{H}}t_{1y} & {\cal{H}}t_{2y}  \\
    0 & 0 & -{\cal{G}} & {\cal{H}}n_z & 0 & 0 & 0 & {\cal{H}}t_{1z} & {\cal{H}}t_{2z}   \\ 
   -{\cal{G}} x & -{\cal{G}} y & -{\cal{G}} z & -{\cal{G}}+{\cal{H}}(\boldsymbol{r}\cdot \boldsymbol{n}) & 0 & 0  &  0 & {\cal{H}}(\boldsymbol{r}\cdot \boldsymbol{t}_{1})  & {\cal{H}}(\boldsymbol{r}\cdot \boldsymbol{t}_{2}) \\
    0 & 0 & 0 & \mu_{oi} {\cal{H}}_{in}n_x & -{\cal{G}}_{in} & 0 & 0 & {\cal{H}}_{in}t_{1x} & {\cal{H}}_{in}t_{2x}  \\
    0 & 0 & 0 & \mu_{oi} {\cal{H}}_{in}n_y & 0 & -{\cal{G}}_{in} & 0 & {\cal{H}}_{in}t_{1y} & {\cal{H}}_{in}t_{2y}  \\ 
    0 & 0 & 0 & \mu_{oi} {\cal{H}}_{in}n_z & 0 & 0 & -{\cal{G}}_{in} & {\cal{H}}_{in}t_{1z} & {\cal{H}}_{in}t_{2z}  \\
   - \epsilon_{io} t_{1x} & - \epsilon_{io} t_{1y} & - \epsilon_{io} t_{1z} & (\epsilon_{io} - \mu_{oi}) \partial_{t_1} & t_{1x} & t_{1y} & t_{1z} & (1 - \epsilon_{io}) {{\cal{K}}_1} & 0     \\
    -\epsilon_{io} t_{2x} & - \epsilon_{io} t_{2y} & - \epsilon_{io} t_{2z} & (\epsilon_{io} - \mu_{oi}) \partial_{t_2} & t_{2x} & t_{2y} & t_{2z} & 0 & (1 - \epsilon_{io}) {{\cal{K}}_2}
  \end{bmatrix}
   \hspace{-0.2 cm}
  \left[ \begin{array}{c} \partial_n H^{scat}_x \\ \partial_n H^{scat}_y \\ \partial_n H^{scat}_z \\ H^{scat}_n \\ \partial_n H^{tran}_x \\ \partial_n H^{tran}_y \\ \partial_n H^{tran}_z \\ H^{scat}_{t_1} \\ H^{scat}_{t_2}  \end{array} \right] 
 \hspace{-0.1 cm} =  \hspace{-0.1 cm}
   \left[ \begin{array}{c}   0 \\ 0 \\ 0 \\ 0 \\  {\cal{A}}^{'}_x  \\ {\cal{A}}^{'}_y  \\  {\cal{A}}^{'}_z  \\ {\cal{B}}^{'}_1 \\ {\cal{B}}^{'}_2  
 \end{array} \right]$ \\
 \\
     (d) $ \epsilon_{oi} \equiv \epsilon_{out}/\epsilon_{in} \equiv 1/\epsilon_{io}$, $\mu_{io} \equiv \mu_{in}/\mu_{out}$, $\partial_n (\cdot) \equiv \partial (\cdot)/ \partial n  \equiv \boldsymbol{n} \cdot \nabla (\cdot)$, $\partial_{t_j} (\cdot) \equiv \partial (\cdot)/ \partial t_j \equiv  \boldsymbol{t}_j \cdot \nabla (\cdot)$ for $j = 1,2$, \\ \noalign{\smallskip} \noalign{\smallskip}
$\begin{aligned} 
\hspace{2.0cm} {\cal{A}}_\alpha &\equiv  - {\cal{H}}_{in}  \left\{ E_{t_1}^{inc}t_{1\alpha} + E_{t_2}^{inc}t_{2\alpha}  + \epsilon_{oi} E_{n}^{inc}n_{\alpha} \right\}, \qquad \qquad \quad  \qquad  (\alpha = x, y, z)   
\\
\hspace{2.0cm} {\cal{B}}_j  &\equiv  \mu_{io} \boldsymbol{t}_{j} \cdot  \frac {\partial{\boldsymbol{E}^{inc}}} {\partial{n}} - (1-\epsilon_{oi})  \frac {\partial{E_n^{inc}}} {\partial{t_{j}}}  + (1 - \mu_{io}) \boldsymbol{n} \cdot \frac {\partial {\boldsymbol{E}^{inc}}} {\partial{t_j}},   \qquad (j = 1,2),  
\\
\hspace{2.0cm} {\cal{C}}_\alpha &\equiv {\cal{H}}  E^{inc}_\alpha, \qquad   (\alpha = x, y, z)    
\qquad \qquad
{\cal{D}}_j  \equiv  \mu_{io} \boldsymbol{t}_{j} \cdot  \frac {\partial{\boldsymbol{E}^{inc}}} {\partial{n}},   \qquad (j = 1,2)   
\\
\hspace{2.0cm} {\cal{A}}^{'}_\alpha &\equiv  - {\cal{H}}_{in}  \left\{ H_{t_1}^{inc}t_{1\alpha} + H_{t_2}^{inc}t_{2\alpha}  + \mu_{oi} H_{n}^{inc}n_{\alpha} \right\}, \qquad \qquad \quad  \qquad  (\alpha = x, y, z)   
\\
\hspace{2.0cm} {\cal{B}}^{'}_j  &\equiv  \epsilon_{io} \boldsymbol{t}_{j} \cdot  \frac {\partial{\boldsymbol{H}^{inc}}} {\partial{n}} - (1-\mu_{oi})  \frac {\partial{H_n^{inc}}} {\partial{t_{j}}}  + (1 - \epsilon_{io}) \boldsymbol{n} \cdot \frac {\partial {\boldsymbol{H}^{inc}}} {\partial{t_j}},   \qquad (j = 1,2).  
\end{aligned}$
\\ \noalign{\smallskip}
        \bottomrule
    \end{tabular}
    \end{ruledtabular}
\end{table*}

Similar remarks apply to the linear system in Table~\ref{Tbl:ls}b except line 4 now represents instead the condition $\nabla \cdot \boldsymbol{E}^{tran} = 0$. Since the combination of Maxwell's equations and the continuity condition of the tangential components of $\boldsymbol{E}$ and $\boldsymbol{H}$ imply the continuity of the normal components of $\boldsymbol{D}$ and $\boldsymbol{B}$, only the divergence free condition in the field on one side of the boundary appear in the linear system in Table~\ref{Tbl:ls}~\cite{StrattonChu_1939}. 
 
Indeed, the systems of equations in Table~\ref{Tbl:ls}a and \ref{Tbl:ls}b can be regarded as the generalized Fresnel equations and Snell's law relating the scattered and transmitted electric fields to the incident electric field at a dielectric interface with prescribed curvature, without any limitation on the magnitude of the curvature relative to the wave number.  In contrast, the familiar Fresnel results are applicable only to a planar interface with zero curvature. In the next subsection we will show how these results simplify for scattering by perfect electrical conductors of general shape as given in (\ref{eq:HG_4N_by_4N}) and (\ref{eq:HG_5N_by_5N}) and how our results reduce to the familiar Fresnel equations and Snell's Law at planar dielectric interfaces.

The presence of numerous zeros in the matrices in (\ref{eq:HG_4N_by_4N}), (\ref{eq:HG_5N_by_5N}), in the linear systems in Tables~\ref{Tbl:ls}a, \ref{Tbl:ls}b and \ref{Tbl:ls}c suggests that the number of equations can be reduced at the expense of more complex matrix coefficients and the unknowns. However, we shall not pursue this simplification here as we focus on our field-only formulation of scattering and transmission.
 
\subsection{\label{sec:Special_cases} Reduction to special cases}
\subsubsection{\label{sec:Recover_PEC} Perfect Electrical Conductor}

The result for perfect electrical conductors (PEC) in (\ref{eq:HG_4N_by_4N}) for the scattered $\boldsymbol{E}$ field follows from Table~\ref{Tbl:ls}a by noting that the transmitted field, $\boldsymbol{E}^{tran} = \boldsymbol{0}$ in this case and the tangential components of the scattered and incident fields cancel: $\boldsymbol{E}_t^{scat} = - \boldsymbol{E}_t^{inc}$. Consequently, the first 4 lines of Table~\ref{Tbl:ls}a reduce to (\ref{eq:HG_4N_by_4N}) and the remaining lines are trivial.

To recover the equation for the scattered $\boldsymbol{H}$ field by a PEC given in (\ref{eq:HG_5N_by_5N}) we can start with the $9N \times 9N$ linear system given in Table~\ref{Tbl:ls}c. The PEC equation for $\boldsymbol{H}$ in (\ref{eq:HG_5N_by_5N}) can now be obtained by taking the limit $\epsilon_{io} \equiv \epsilon_{in}/\epsilon_{out} \rightarrow \infty$, for finite $\mu_{oi}$. In this limit, the transmitted field, $\boldsymbol{H}^{tran}$ vanishes, and (\ref{eq:div_EH_b}), namely, $\nabla \cdot \boldsymbol{H} = 0$ is satisfied on the PEC surface. This means that the 4th line of Table~\ref{Tbl:ls}c is satisfied automatically. We saw earlier that this is also equivalent to (\ref{eq:Hn_eq_0}), that is, the normal component of the scattered field cancels that of the incident field, so we can replace the unknown $ H_n^{scat}$ in Table~\ref{Tbl:ls}c by $(-H_n^{inc})$ that is known, and thus we are left with only 5 equations from lines 1 to 3 and lines 8 and 9 in Table~\ref{Tbl:ls}c. And finally, to obtain the same equations as in the PEC result (\ref{eq:HG_5N_by_5N}), the terms ${\cal{B}}^{'}_j$ in Table~\ref{Tbl:ls}d reduce to ${\cal{Z}}_j$ in (\ref{eq:HG_5N_by_5N}) when we use the following relations from differential geometry between the surface normal, $\boldsymbol{n}$, surface tangents, $\boldsymbol{t}_j$ and local curvature, $\kappa_j$: $\partial \boldsymbol{n} / \partial \boldsymbol{t}_j = \kappa_j \boldsymbol{t}_j$, $\partial \boldsymbol{t}_j / \partial \boldsymbol{t}_j = -\kappa_j \boldsymbol{n}, (j = 1, 2)$ and $\partial \boldsymbol{t}_1 / \partial \boldsymbol{t}_2 = \boldsymbol{0}$.


\subsubsection{\label{sec:Recover_Fresnel} Planar Dielectric - Fresnel equations and Snell's law}

We now show how to recover from our field-only formulation in Tables~\ref{Tbl:ls}a and \ref{Tbl:ls}b, the Fresnel equations and Snell's law that describe scattering and transmission of the $\boldsymbol{E}$ field across a \emph{planar} interface located at $z=0$. 

Consider the scattering of an incident $s$-polarized or transverse-electric (TE) plane wave given by the incident electric field and incident wave vector
\begin{subequations}  \label{eq:plane_wave_E_inc}
  \begin{eqnarray}
 \boldsymbol{E}^{inc} &=& E^{inc}_0 (0,\exp[i k_{out}(- x \sin\theta_{i}  + z \cos\theta_{i})],0) \qquad \label{eq:plane_wave_E_inc_a}
  \\
 \boldsymbol{k}_{out} &=& k_{out} (-\sin\theta_{i}, 0, \cos\theta_{i}).  \label{eq:plane_wave_E_inc_b}
\end{eqnarray}
\end{subequations}
The outward surface normal is $\boldsymbol{n} = (0,0,1)$ and the angle of incidence, $\theta_i$, measured relative to $\boldsymbol{n}$, is given by $k_{out} \cos \theta_i = \boldsymbol{k}_{out} \cdot \boldsymbol{n}$. We take as the two tangential unit vectors: $\boldsymbol{t}_1 = \boldsymbol{t}_y = (0,1,0)$ and $\boldsymbol{t}_2 = \boldsymbol{t}_x = (1,0,0)$ and note that the curvatures, $\kappa_j$ are zero for a planar interface. For this geometric configuration, Table~\ref{Tbl:ls}b for the transmitted electric field can now be simplified using the above definitions of the surface normal and tangents and the explicit form for $\boldsymbol{E}^{inc}$ to give

 \begin{widetext}
\begin{align} \label{eq:HG_9N_by_9N_tran_Fresnel_1}
  \hspace{-0.4 cm}
  \begin{bmatrix}
   -{\cal{G}} & 0 & 0  & 0 & 0 & 0 & 0 & {\cal{H}} & 0  \\
    0 & -{\cal{G}} & 0 & 0 & 0 & 0 & 0 & 0 & {\cal{H}}  \\
    0 & 0 & -{\cal{G}} & \epsilon_{io} {\cal{H}} & 0 & 0 & 0 & 0 & 0   \\ 
   0 & 0 & 0 &  -{\cal{G}}_{in} & -{\cal{G}}_{in} x & -{\cal{G}}_{in} y & 0 & {\cal{H}}_{in} x  & {\cal{H}}_{in} y \\
    0 & 0 & 0 & 0 & -{\cal{G}}_{in} & 0 & 0 & {\cal{H}}_{in} & 0  \\
    0 & 0 & 0 & 0 & 0 & -{\cal{G}}_{in} & 0 & 0 & {\cal{H}}_{in}  \\ 
    0 & 0 & 0 & {\cal{H}}_{in} & 0 & 0 & -{\cal{G}}_{in} & 0 & 0  \\
   - \mu_{io} & 0 & 0 & ( \epsilon_{oi} \mu_{io} - 1) \partial_{x} & 1 & 0 & 0 & 0 & 0     \\
    0 & - \mu_{io} & 0 & 0 & 0 & 1 & 0 & 0 & 0
  \end{bmatrix}
   \hspace{-0.2 cm}
  \left[ \begin{array}{c} \partial_z E^{scat}_x \\ \partial_z E^{scat}_y \\ \partial_z E^{scat}_z \\ E^{tran}_z \\ \partial_z E^{tran}_x \\ \partial_z E^{tran}_y \\ \partial_z E^{tran}_z \\ E^{tran}_{x} \\ E^{tran}_{y}  \end{array} \right] 
 \hspace{-0.1 cm} =  \hspace{-0.1   cm}
  & \left[ \begin{array}{c}   0 \\ {\cal{H}} E^{inc}_y \\ 0 \\ 0 \\  0  \\ 0  \\  0  \\ 0 \\ \mu_{io} (\partial E^{inc}_y / \partial z)
 \end{array} \right].
\end{align}
\end{widetext}

From (\ref{eq:HG_9N_by_9N_tran_Fresnel_1}), it follows that the following quantities: $\partial E^{scat}_x / \partial z$, $\partial E^{scat}_z / \partial z$, $E^{tran}_z$, $\partial E^{tran}_x / \partial z$, $\partial E^{tran}_z / \partial z$ and $E^{tran}_x$ all vanish and the remaining three unknowns: $\partial E^{scat}_y / \partial z$, $\partial E^{tran}_y / \partial z$ and $E^{tran}_y$, satisfy the equations

\begin{subequations}  \label{eq:Three_eqns_Fresnel_1}
\begin{eqnarray}
-{\cal{G}} \frac{\partial E^{scat}_y}{\partial z} + {\cal{H}} E^{tran}_y &=&   {\cal{H}} E^{inc}_y   \label{eq:Three_eqns_Fresnel_1_a} 
 \\
  -{\cal{G}}_{in} \frac{\partial E^{tran}_y}{\partial z} + {\cal{H}}_{in} E^{tran}_y &=&   0    \label{eq:Three_eqns_Fresnel_1_b}
  \\
    - \mu_{io} \frac{\partial E^{scat}_y}{\partial z} +  \frac{\partial E^{tran}_y}{\partial z}&=&   \mu_{io} \frac{\partial E^{inc}_y}{\partial z}  \label{eq:Three_eqns_Fresnel_1_c}
\end{eqnarray}
\end{subequations}

The continuity of the tangential component of $\boldsymbol{E}$ implies at the interface, $z=0$
\begin{eqnarray} 
  E^{inc}_y + E^{scat}_y = E^{tran}_y.
\end{eqnarray}
so on combining this with (\ref{eq:Three_eqns_Fresnel_1_a}) and (\ref{eq:Three_eqns_Fresnel_1_b}) we find 
\begin{subequations}  \label{eq:scat_tran_Fresnel_1}
\begin{eqnarray}
{\cal{G}} \frac{\partial E^{scat}_y}{\partial z} &=&   {\cal{H}} E^{scat}_y   \label{eq:scat_tran_Fresnel_1_a} 
 \\
 {\cal{G}} \frac{\partial E^{tran}_y}{\partial z} &=&   {\cal{H}} E^{tran}_y.       \label{eq:scat_tran_Fresnel_1_b}
\end{eqnarray}
\end{subequations}

At a planar interface, the surface integrals: ${\cal{G}}$, ${\cal{H}}$, ${\cal{G}}_{in}$, ${\cal{H}}_{in}$ are independent of $\boldsymbol{r}_0$ so the solution can be represented as
\begin{subequations} \label{eq:plane_wave_solutions}
\begin{eqnarray}
  E^{inc}_y &=&  E^{inc}_0 \exp [i k_{out} (-x \sin \theta_{i} + z \cos \theta_{i}) ]
 \\
   E^{scat}_y &=&  E^{scat}_0 \exp [i k_{out} (-x \sin \theta_{i} - z \cos \theta_{i}) ]
   \\
    E^{tran}_y &=&  E^{tran}_0 \exp [i k_{in} (-x \sin \theta_{t} + z \cos \theta_{t}) ]
\end{eqnarray}
\end{subequations}
with $\theta_t$ given by $k_{in}\cos \theta_t = \boldsymbol{k}_{in} \cdot \boldsymbol{n}$. Snell's law then follows from matching of the phase factor at $z=0$:
\begin{eqnarray} \label{eq:snell}
  k_{out} \sin \theta_{i} = k_{in} \sin \theta_{t}.
\end{eqnarray}

By combining (\ref{eq:Three_eqns_Fresnel_1_c}), (\ref{eq:plane_wave_solutions}) and (\ref{eq:snell}) we obtain the well-known Fresnel formula relating the scattered field amplitude to the incident field amplitude~\cite{Slater_Frank_1947}

\begin{subequations} \label{eq:fresnel_ratio_1}
\begin{eqnarray}
 \frac{E^{scat}_0} {E^{inc}_0} &=& \frac {\mu_{in} \tan \theta_t - \mu_{out} \tan \theta_i} {\mu_{in} \tan \theta_t + \mu_{out} \tan \theta_i}
  \\
  &=&  \frac {\sin (\theta_t - \theta_i) } {\sin (\theta_t + \theta_i)}, \quad \text{if} \; \;\mu_{in}=\mu_{out}.
\end{eqnarray}
\end{subequations}

For the scattering of an incident $p$-polarized or transverse-magnetic (TM) plane wave given by the incident electric field

 \begin{eqnarray} \label{eq:plane_wave_E_inc_TM}
 \boldsymbol{E}^{inc} =E^{inc}_0  \left[ \begin{array} {c} \cos \theta_i \\ 0 \\ \sin \theta_i\end{array} \right]  \exp[i k_{out}(- x \sin\theta_{i}  + z \cos\theta_{i}). \qquad \;
\end{eqnarray}
The matrix for the linear system is the same as that in (\ref{eq:HG_9N_by_9N_tran_Fresnel_1}) for the $s$-polarized TE incident wave. Thus after some algebra, we obtain Snell's law and the known Fresnel result\cite{Slater_Frank_1947}

\begin{subequations} \label{eq:fresnel_ratio_TM}
\begin{eqnarray}
 \frac{E^{scat}_0} {E^{inc}_0} &=& \frac {\mu_{in} \sin \theta_t \cos \theta_t - \mu_{out} \sin \theta_i \cos \theta_i} {\mu_{in} \sin \theta_t \cos \theta_t + \mu_{out} \sin \theta_i \cos \theta_i}
  \\
  &=&  \frac {\tan (\theta_t - \theta_i) } {\tan (\theta_t + \theta_i)}, \quad \text{if} \; \;\mu_{in}=\mu_{out}.
\end{eqnarray}
\end{subequations}

It can be concluded that the linear systems in Table~\ref{Tbl:ls} embody the boundary integral generalizations of the Fresnel equations and Snell's law at a curved interface for the scattering and transmission of $\boldsymbol{E}$ and $\boldsymbol{H}$.

\subsection{\label{sec:DIEL_results} Dielectric results}
\subsubsection{\label{sec:Diel_Mie_sphere} Dielectric Sphere - Mie}

\begin{figure}[t]
\centering{}\includegraphics[width=3in]{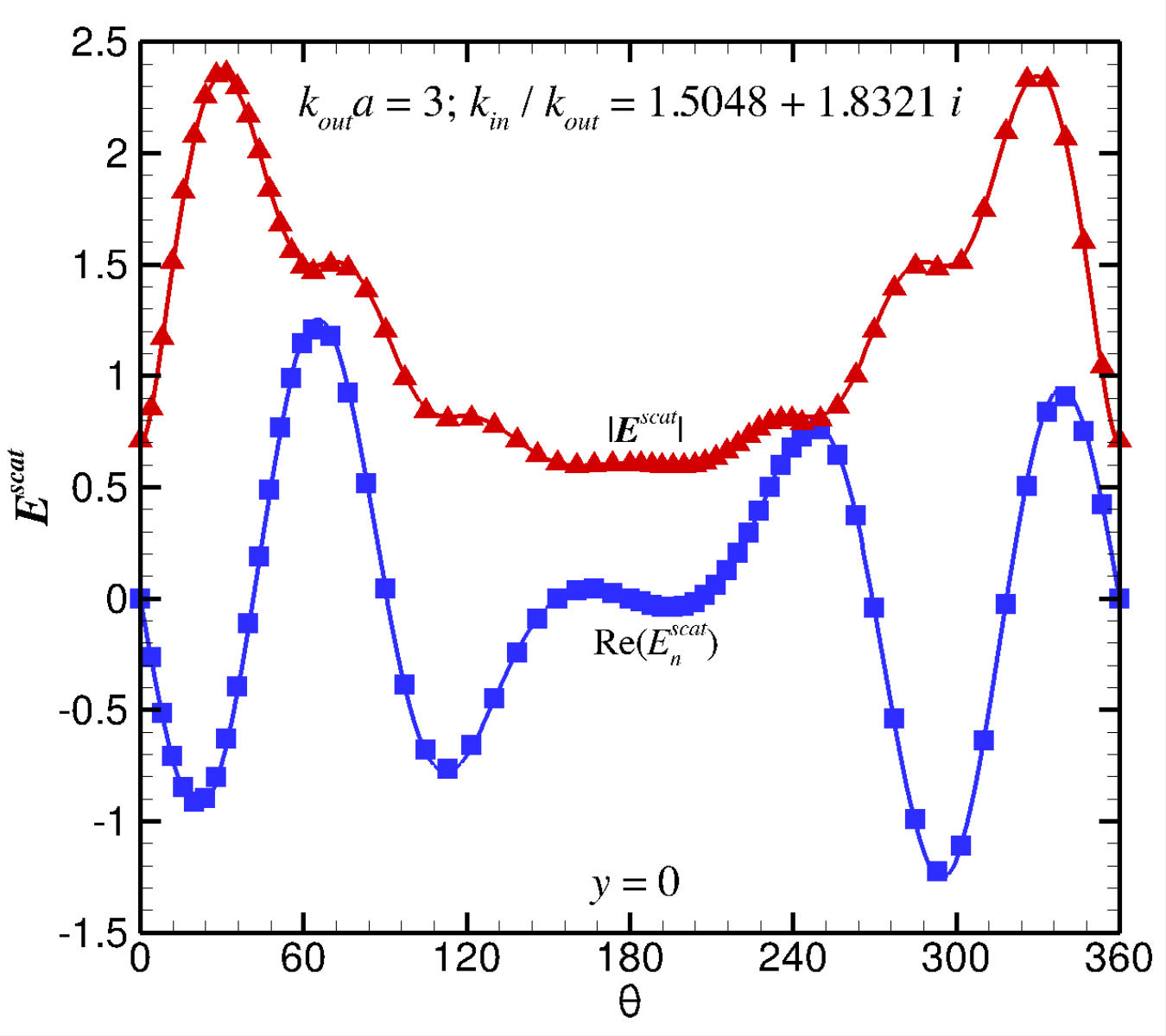} \caption{\label{Fig:Dielectric_Mie_E} (Color on-line)  Variation of the scattered field: the real part of the normal component, $E_n^{scat}$ and its magnitude, $|\boldsymbol{E}^{scat}|$ on the surface of a dielectric sphere in an incident field $\boldsymbol{E}^{inc} = (1,0,0) \exp (ik_{out}z)$, at $k_{out}a = 3$, $k_{in}/k_{out}=1.5048+1.8321~i$ along the meridian at $y=0$. Using $N=2562$ nodes and 1280 quadratic elements, the absolute difference between the present field-only approach (symbols) and the analytical Mie theory (lines) is less than 0.02.}
\end{figure}

In Fig.~\ref{Fig:Dielectric_Mie_E} we show numerical results for the scattered field on the surface of a dielectric sphere (radius $a$) subject to an incident plane wave $\boldsymbol{E}^{inc} = (1,0,0) \exp(ik_{out}z)$ at $k_{out}a = 3$. The dielectric sphere has a constant but complex relative permittivity that corresponds to a 200~nm radius gold nano sphere at wavelength of 418.9 nm \cite{Rakic_1998} in air, whereby $k_{in}/k_{out} = (\epsilon_{in}/\epsilon_{out})^{1/2} =1.5048+1.8321~i$. 

We compare results from our field-only PEC-E approach with the analytic results from the Mie theory.  Since the dielectric permittivity of the sphere is complex, we show results for the magnitude of the scattered field, $|\boldsymbol{E}^{scat}|$ and the real part of the normal component of the scattered field, $\text{Re}(E_n^{scat})$. This example demonstrates the flexibility of the present formulation in being able to handle propagation in media with complex dielectric permittivities.

\begin{figure} [ht]
  \centering{}
  \subfloat[]{ \includegraphics[width=3in]{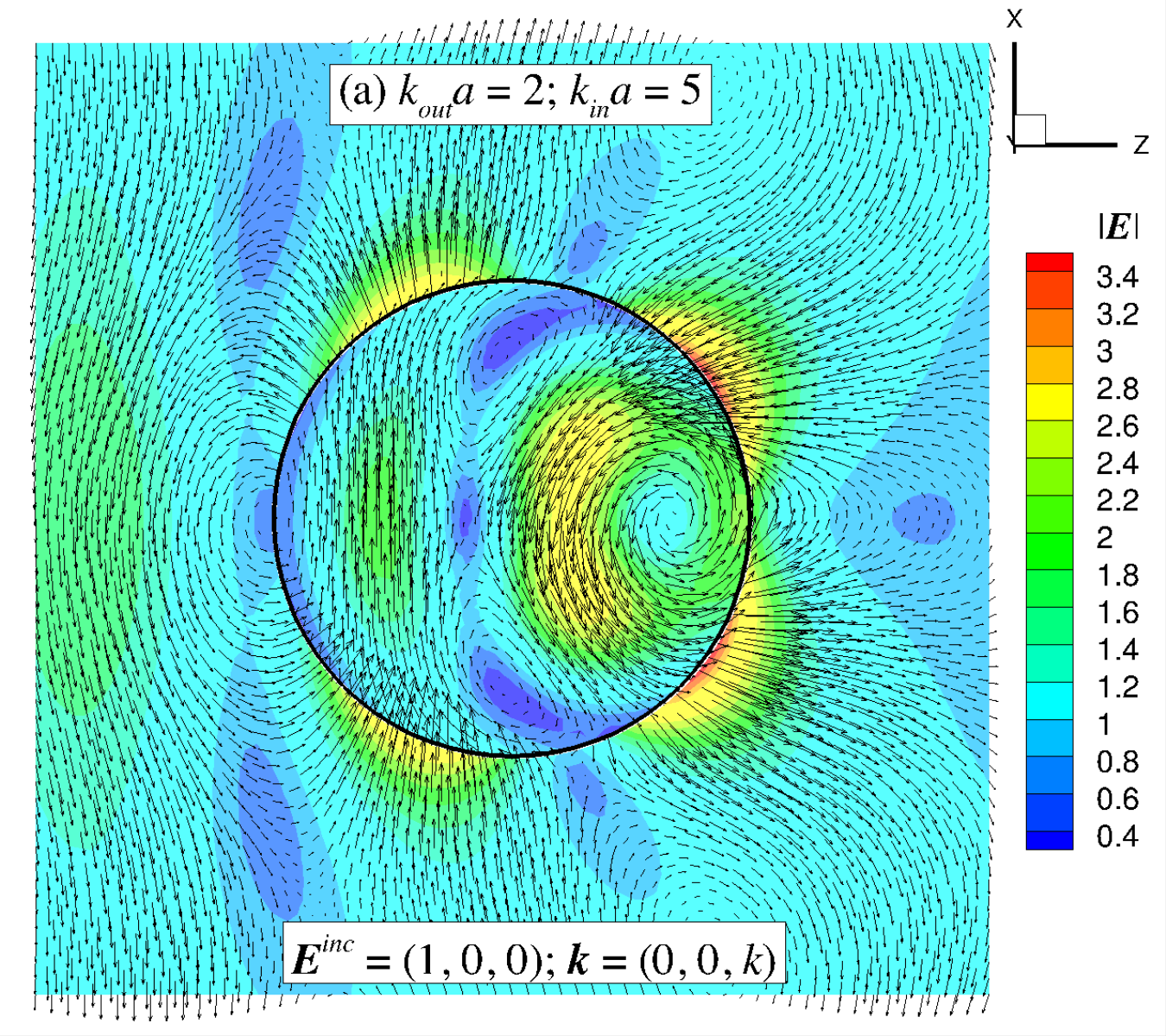} }\\
  \subfloat[]{ \includegraphics[width=3in]{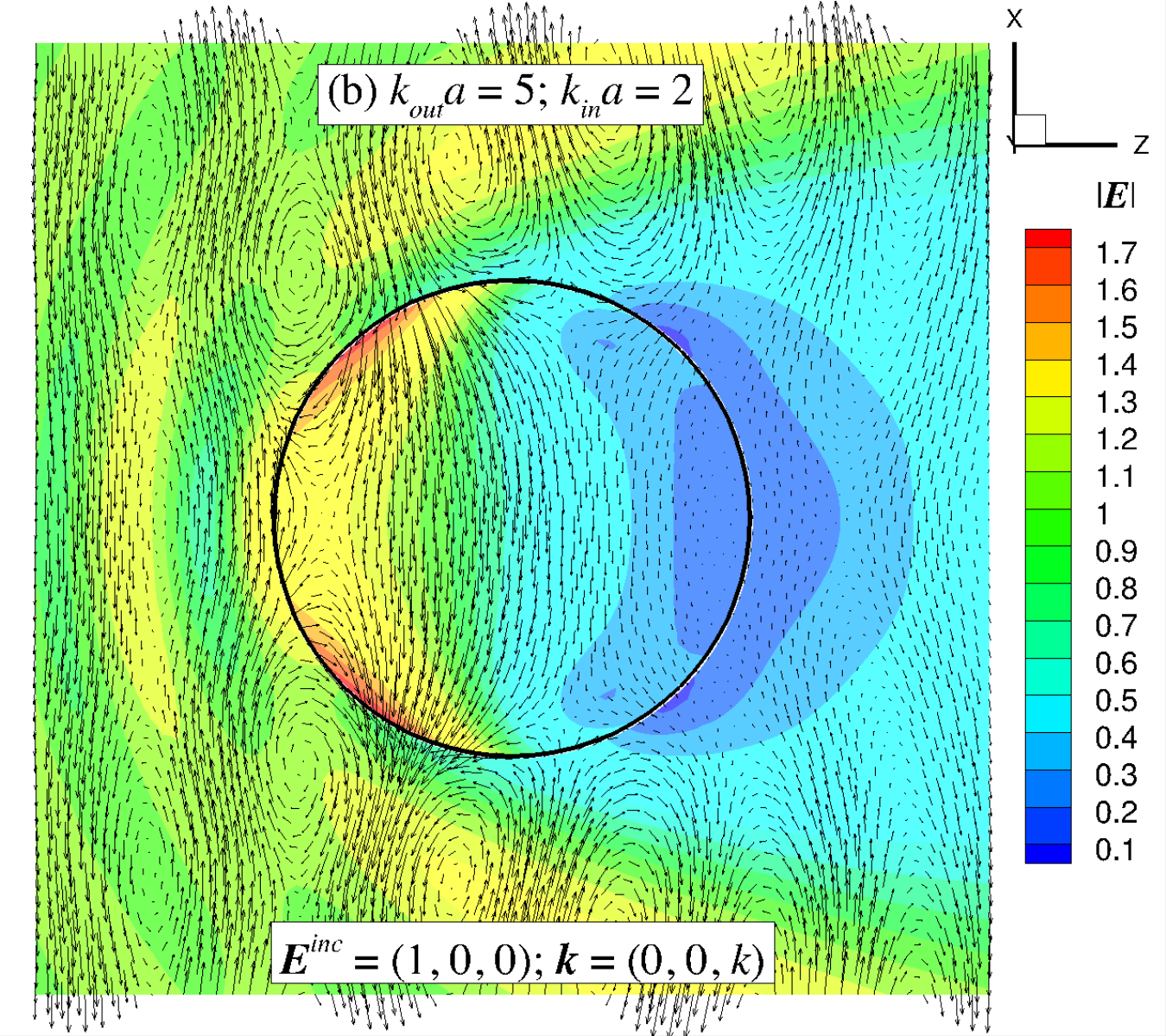} } 
\caption{\label{Fig:Dielectricsphere}  (Color on-line) The total field, $\boldsymbol{E}$ in the plane $y=0$ around and inside a dielectric sphere due to an incident electric field, $\boldsymbol{E}^{inc} = (1,0,0) \exp (ikz)$ with (a) $k_{out}a = 2 $, $k_{in}a= 5$ and (b) $k_{out}a=5$, $k_{in}a=2$.  Results are obtained using 1442 nodes and 720 quadratic elements sphere. (Animation on-line)} 
\end{figure}

In Fig.~\ref{Fig:Dielectricsphere}, we show vector plots of the total field, $\boldsymbol{E}$ in the plane $y=0$ around a dielectric sphere subject to the same plane wave at (a) $k_{out}a = 2.0$ and $k_{in}a = 5.0$ and (b) $k_{out}a = 5.0$ and $k_{in}a = 2.0$. The very different diffraction effects for the two cases are evident.

In the case of Fig.~\ref{Fig:Dielectricsphere}a for $k_{out}a = 2.0$ and $k_{in}a = 5.0$, the incoming wave generates a vortex-alike structure in the internal field located on the downstream half of the sphere - this is even more apparent in the accompanying movie-file, see supplementary material. In the upstream portion of the sphere, the internal electric field is aligned more or less along the $x$-direction. In between these two regions, around the center of the sphere, the absolute value of the electric field exhibits three separate minima, and the electric field has the largest magnitude on the downstream exterior surface. A few local minima can also be observed on the upstream exterior of the sphere.

In contrast, the field structure is quite different in Fig.~\ref{Fig:Dielectricsphere}b, where we have interchanged the $ka$ values from $k_{out}a=5.0$ and $k_{in}a=2.0$ to $k_{out}a=2.0$ and $k_{in}a=5.0$. Now, the magnitude of the electric field assumes a maximum in the upstream part of the sphere. The incoming wave is being scattered to the sides by the dielectric sphere - again, this effect is most clearly visible in the movie-file of the supplementary material. The result then is a large shadow region on the downstream side of the sphere with a small field magnitude. This is accompanied by an envelope of higher field intensity (``yellow'' color) that is due to the constructive interference between the incident and reflected wave.
 

\subsubsection{\label{sec:Layered_Diel_spheres} Layered Dielectric Spheres}

\begin{figure} [ht]
  \centering{}
  \subfloat[]{ \includegraphics[width=3in]{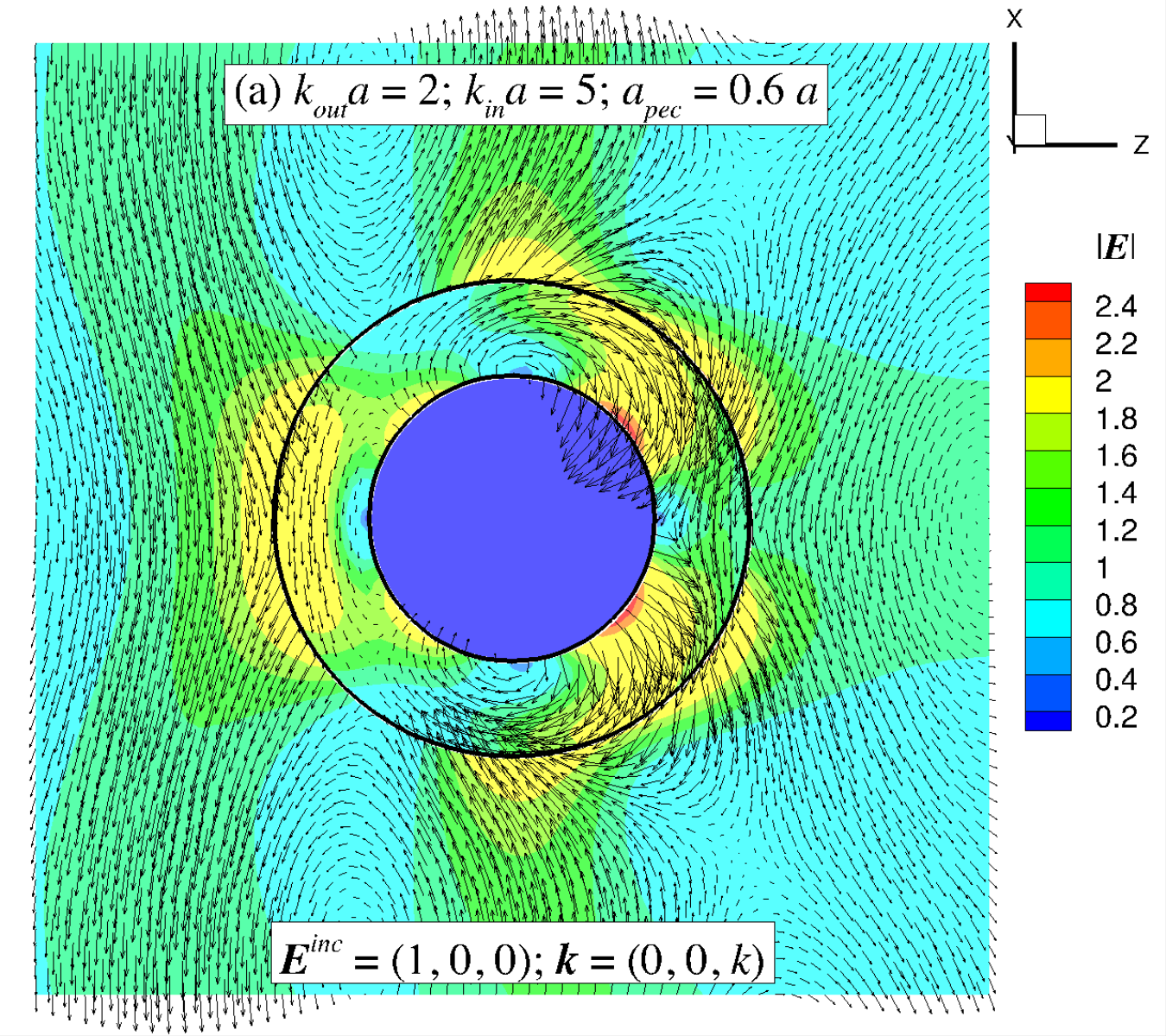} }\\
  \subfloat[]{ \includegraphics[width=3in]{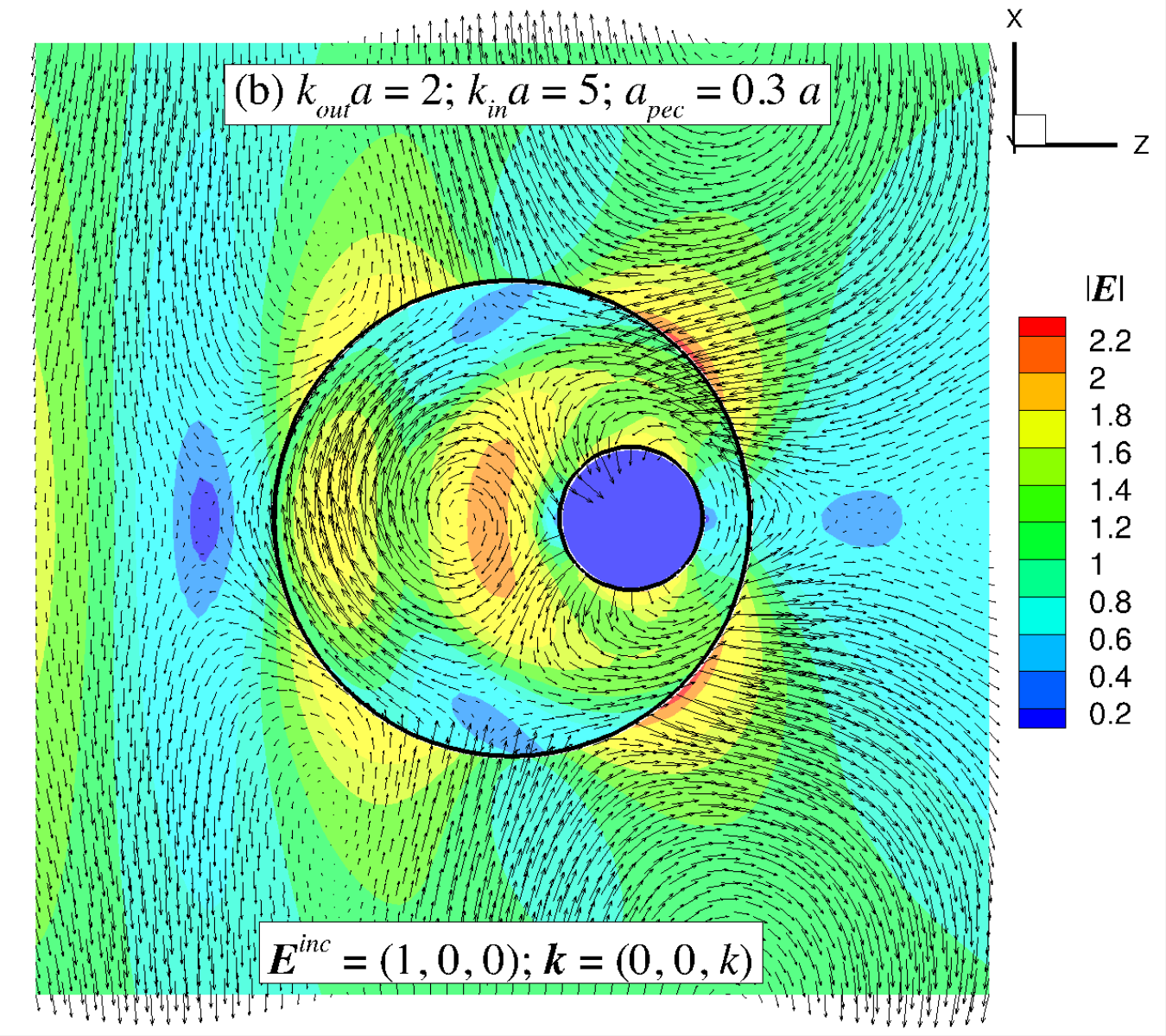} } 
\caption{\label{Fig:Dielectriclayersphere}  (Color on-line) The total field, $\boldsymbol{E}$ in the plane $y=0$ around a dielectric sphere with a PEC sphere inside it due to an incident electric field, $\boldsymbol{E}^{inc} = (1,0,0) \exp (ikz)$ with $k_{out}a = 2 $ and $k_{in}a= 5$ for (a) a concentric PEC sphere of radius $a_{pec} = 0.6 a$; and (b) a PEC sphere of radius $a_{pec} = 0.3 a$ whose centre is at $a/2$ from the centre of the dielectric sphere.  Results are obtained using 1442 nodes and 720 quadratic elements on each sphere. (Animation on-line)} 
\end{figure}

Based on these observations, it would be interesting to see what happens if a small PEC sphere is placed inside the dielectric sphere shown in Fig.~\ref{Fig:Dielectricsphere}a with $k_{out}a=2.0$ and $k_{in}a=5.0$. First, we embed a concentric PEC sphere, with radius 0.6 of that of the dielectric sphere and the resulting vector plot of the total field, $\boldsymbol{E}$ in the plane $y=0$ is shown in Fig.~\ref{Fig:Dielectriclayersphere}a. When compared to Fig.~\ref{Fig:Dielectricsphere}a, we see that both the spatial structure and magnitudes of the electric field have changed drastically. Regions that previously have low electric field strengths now have large field strengths. The largest values of the electric field occur at the downstream surface of the PEC sphere, but its absolute value, $|\boldsymbol{E}| \sim 2.4$, is smaller than the maximum value observed in Fig.~\ref{Fig:Dielectricsphere}a, $|\boldsymbol{E}| \sim 3.4$. The observed effects are mainly caused by the fact that the electric field is forced to be perpendicular to the surface of the embedded PEC sphere. 

In order to investigate further the influence of the presence of an embedded PEC object, we place a smaller PEC sphere, 0.3 times the radius of the dielectric sphere, midway between the center and surface of the dielectric sphere, see Fig.~\ref{Fig:Dielectriclayersphere}b where the interior field of the dielectric sphere has a maximum adjacent to a vortical structure, see Fig.~\ref{Fig:Dielectricsphere}a. The result in Fig.~\ref{Fig:Dielectriclayersphere}b clearly shows that the presence of a relatively small PEC sphere can have a large influence on the behavior of the electric field both inside and outside the dielectric sphere. The vortical field structure in the internal field is still present, but is clearly modified by the presence of the small PEC inclusion (see also the movie-file in the supplementary material). The maximum total electric field magnitude has been reduced to $|\boldsymbol{E}| \sim 2.2$.

\section{\label{sec:ZEROFREQ} Low frequency behavior}

To demonstrate that our field-only formulation is numerically robust in the long wavelength limit we solve the PEC-E equations for spheres with $ka << 1$ and compare with known analytical results for $ka \equiv 0$. We consider a dielectric sphere (radius $a$) of permittivity, $\epsilon_{in}$ in an external medium, $\epsilon_{out}$ with a concentric spherical cavity inclusion of radius, $a_{cav} < a$ and permittivity, $\epsilon_{cav}$. When this layered sphere is placed in an electrostatic incident field, $\boldsymbol{E}_{inc} = (E_0,0,0)$ the potential, $V$ has the solution
\begin{subequations} \label{eq:dielec_shell_potential}
\begin{eqnarray}
  V &=&  - E_0 \; r \cos  \theta + A  (a^3/r^2) \cos  \theta, \qquad   r > a
\\
    &=&  B \; r \cos  \theta + C (a^3/r^2) \cos  \theta, \; a_{cav} < r < a
\\
   &=&  -E_{cav} \; r \cos  \theta = -E_{cav} \; x,  \qquad   \quad  r < a_{cav}.
\end{eqnarray}
\end{subequations}
\begin{figure}[ht]
\centering{}\includegraphics[width=3in]{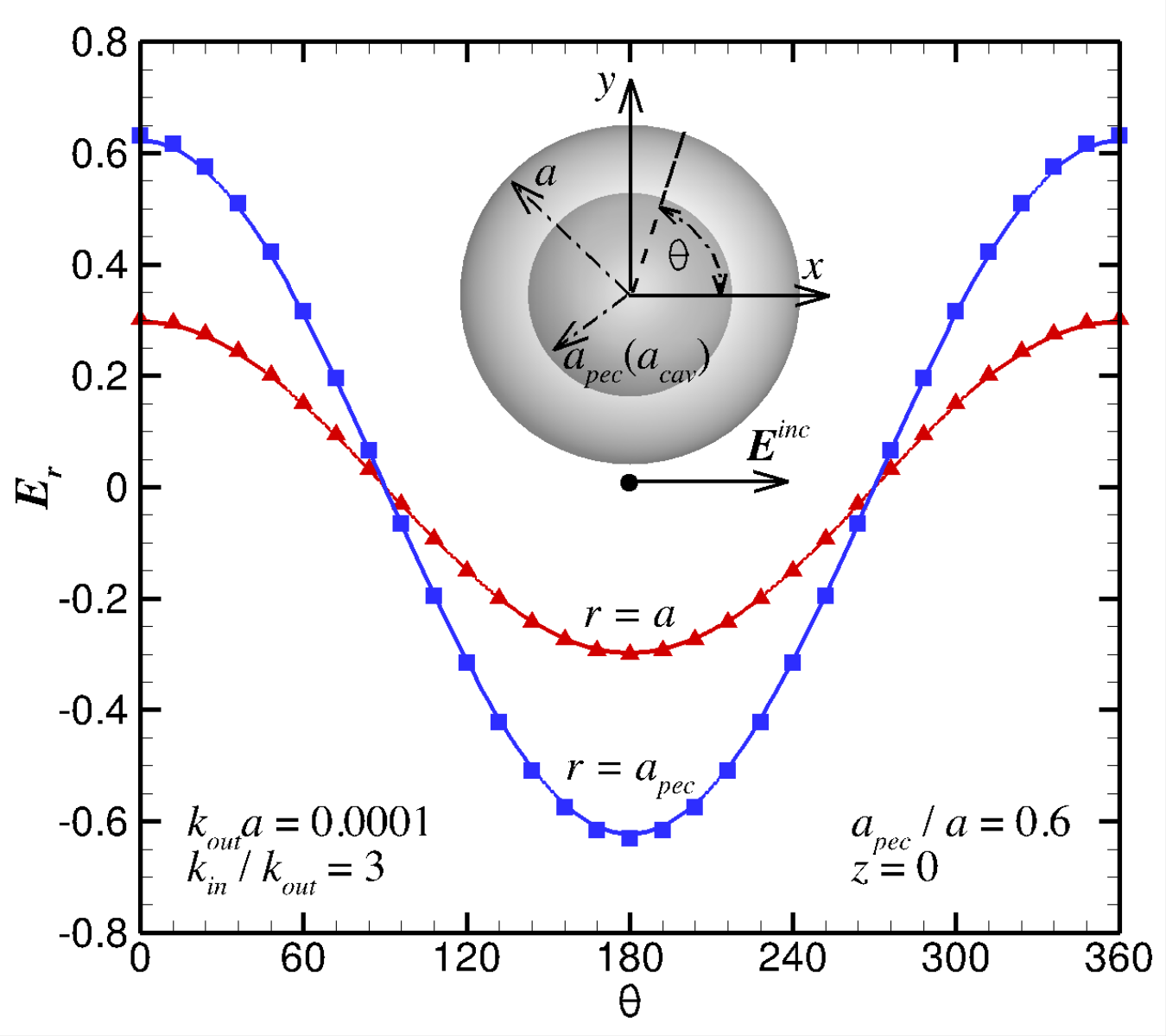} \caption{\label{Fig:ksmall} (Color on-line) The radial component of electric fields $E_{r}$ at $r = a$ and at $r = a_{pec}$ for a dielectric sphere with (radius, $a$) with a concentric perfect electrical conducting core (radius, $a_{pec} = 0.6a$) in an incident field $E^{inc}= (1, 0, 0) \exp (ik_{out}z)$ at $k_{out}a = 0.0001$. Using 1442 nodes and 720 quadratic elements the absolute difference from the analytical (lines) and numerical (symbols) solutions is less than 0.01.}
\end{figure}
The constants $A, B, C$ and the constant cavity field magnitude, $E_{cav}$~\cite{LandauLifshitz_1960}, can be found from the continuity conditions of $V$ and $\epsilon (\partial V/\partial r)$ at $r = a$ and $a_{cav}$ and the field is then given by $\boldsymbol{E}=-\nabla V$.

In Fig.~\ref{Fig:ksmall}, we show results for the radial component of the electric field, $E_r$  at $k_{out}a= 0.0001$, $k_{in}/k_{out} = 3$ for a concentric spherical PEC inclusion ($\epsilon_{cav} \rightarrow \infty$) and $a_{cav} \equiv a_{pec} = 0.6 a$. The absolute difference between our field-only approach and the analytic result (\ref{eq:dielec_shell_potential}) is less than 0.01. This therefore demonstrates that our field-only formulation is numerically robust at all frequencies or wavelengths and thus provides a simple solution for the zero frequency catastrophe.

\section{\label{sec:CONCLUDE} Conclusions}

We have developed a formulation for the numerical solution of the Maxwell's equations in the frequency domain in piecewise homogeneous materials characterized by linear constitutive equations. The key feature of the approach is that one can obtain the electric field directly by solving scalar Helmholtz equations for the Cartesian components of $\boldsymbol{E}$ and the scalar function ($\boldsymbol{r \cdot E}$). The magnetic field can be obtained independently by solving the analogous scalar Helmholtz equations for the Cartesian components of $\boldsymbol{H}$ and the scalar function ($\boldsymbol{r \cdot H}$). As such, conventional boundary integral methods for solving the scalar Helmholtz equation can be used.

The implementation of this formalism is made more robust numerically by removing analytically all singularities that arise from the Green's function and the term involving the solid angle that appear in the conventional boundary integral method. Such a formulation facilitates the use of higher order surface elements that can represent the surface geometry more accurately as well as the use of accurate quadrature methods to compute the surface integrals.

At a more fundamental level, our field-only formulation has also removed the troublesome zero frequency catastrophe that causes the widely adopted surface currents approach to fail numerically in the long wavelength limit. In practical terms, our formulation has done away with having to handle principal value surface integrals in which the inherent divergent behavior precludes the accurate evaluation of field values near boundaries and causes loss of precision in geometries where the separation between two surfaces is small compared to the characteristic wavelength.

It is well known that spurious resonant solutions can appear in numerical solutions of the boundary integral formulation of the Helmholtz equation if the wave number, $k$ is close to one of the eigenvalues of the problem. Our initial investigations suggest that with our non-singular formulation of the boundary integral equation (BRIEF), the value of $k$ has to be within 0.1\% of an eigenvalue before the effects of the resonant solution can be significant~\cite{Sun_2015b}. However, it remains an open problem as to how to ameliorate this issue in practice or to exploit this as a way to find such resonant frequencies that are important in surface plasmonics.  

As mentioned at the end of Sec. \ref{sec:DIEL_formulation}, it is possible to take advantage the large number of zero entries in the linear system to reduce the size of matrix equations. This is a direction that is worthy of further development. 

For our examples, that are relatively simple, we use Gauss elimination to solve the linear system. For large and complex problems iterative solvers or faster $(N \log N)$ algorithms can also be adopted.

\begin{acknowledgments}
This work is supported in part by the Australian Research Council through a Discovery Early Career Researcher Award to QS and a Discovery Project Grant to DYCC.
\end{acknowledgments}

\nocite{*}


\providecommand{\noopsort}[1]{}\providecommand{\singleletter}[1]{#1}%

\end{document}